\documentclass[preprint]{aastex}      

\shorttitle{SDSS Quasar Catalog III.}  
\shortauthors{Schneider et al.}        

\begin{document}

\title{The Sloan Digital Sky Survey Quasar Catalog~III. Third Data Release
}

\author{
Donald~P.~Schneider,\altaffilmark{1}
Patrick~B.~Hall,\altaffilmark{2,3}
Gordon~T.~Richards,\altaffilmark{3}
Daniel~E.~Vanden~Berk,\altaffilmark{1}
Scott~F.~Anderson,\altaffilmark{4}
Xiaohui~Fan,\altaffilmark{5}
Sebastian~Jester,\altaffilmark{6}
Chris~Stoughton,\altaffilmark{6}
Michael~A.~Strauss,\altaffilmark{3}
Mark~SubbaRao,\altaffilmark{7,8}
W.N.~Brandt,\altaffilmark{1}
James~E.~Gunn,\altaffilmark{3}
Brian~Yanny,\altaffilmark{6}
Neta~A.~Bahcall,\altaffilmark{3}
J.C.~Barentine,\altaffilmark{9}
Michael~R.~Blanton,\altaffilmark{10}
William~N.~Boroski,\altaffilmark{6}
Howard~J.~Brewington,\altaffilmark{9}
J.~Brinkmann,\altaffilmark{9}
Robert~Brunner,\altaffilmark{11}
Istv\'an~Csabai,\altaffilmark{12}
Mamoru~Doi,\altaffilmark{13}
Daniel~J.~Eisenstein,\altaffilmark{5}
Joshua~A.~Frieman,\altaffilmark{7}
Masataka~Fukugita,\altaffilmark{14,15}
Jim~Gray,\altaffilmark{16}
Michael~Harvanek,\altaffilmark{9}
Timothy~M.~Heckman,\altaffilmark{17}
\v Zeljko~Ivezi\'c,\altaffilmark{4}
Stephen~Kent,\altaffilmark{6}
S.J.~Kleinman,\altaffilmark{9}
Gillian~R.~Knapp,\altaffilmark{3}
Richard~G.~Kron,\altaffilmark{6,7}
Jurek~Krzesinski,\altaffilmark{9,18}
Daniel~C.~Long,\altaffilmark{9}
Jon~Loveday,\altaffilmark{19}
Robert~H.~Lupton,\altaffilmark{3}
Bruce~Margon,\altaffilmark{20}
Jeffrey~A.~Munn,\altaffilmark{21}
Eric~H.~Neilsen,\altaffilmark{9}
Heidi~Jo~Newberg,\altaffilmark{22}
Peter~R.~Newman,\altaffilmark{9}
R.C.~Nichol,\altaffilmark{23}
Atsuko~Nitta,\altaffilmark{9}
Jeffrey~R.~Pier,\altaffilmark{21}
Constance~M.~Rockosi,\altaffilmark{24}
David~H.~Saxe,\altaffilmark{15}
David~J.~Schlegel,\altaffilmark{3,25}
Stephanie~A.~Snedden,\altaffilmark{9}
Alexander~S.~Szalay,\altaffilmark{17}
Aniruddha~R.~Thakar,\altaffilmark{17}
Alan~Uomoto,\altaffilmark{26}
and
Donald~G.~York\altaffilmark{7,27}
}


\newcounter{address}
\setcounter{address}{1}
\altaffiltext{1}{Department of Astronomy and Astrophysics, The
   Pennsylvania State University, University Park, PA 16802.
   (First Author's email address \hbox{is dps@astro.psu.edu.)}}
\altaffiltext{2}{Department of Physics and Astronomy, York University,
   4700 Keele Street, Toronto, Ontario M3J 1P3, Canada.}
\altaffiltext{3}{Princeton University Observatory, Princeton,
   NJ 08544.}
\altaffiltext{4}{Department of Astronomy, University of Washington,
   Box 351580, Seattle, WA 98195.}
\altaffiltext{5}{Steward Observatory, The University of Arizona,
   933~North Cherry Avenue, Tucson, AZ 85721.}
\altaffiltext{6}{Fermi National Accelerator Laboratory, P.O. Box 500,
   Batavia, IL 60510.}
\altaffiltext{7}{Astronomy and Astrophysics Center, University of
   Chicago, 5640 South Ellis Avenue, Chicago, IL 60637.}
\altaffiltext{8}{Adler Planetarium, Chicago, IL 60605.}
\altaffiltext{9}{Apache Point Observatory, P.O. Box 59,
   Sunspot, NM 88349-0059.}
\altaffiltext{10}{Department of Physics, New York University,
   4 Washington Place, New York, NY 10003.}
\altaffiltext{11}{Department of Astronomy, University of
   Illinois, 1002 W. Green St., Urbana, IL 61801.}
\altaffiltext{12}{Department of Physics of Complex Systems, E\"otv\"os
   Lor\'and University, Pf. 32, H-1518 Budapest, Hungary.}
\altaffiltext{13}{Department of Astronomy and Research Center for the
   Early Universe, School of Science, University of Tokyo, Mitaka,
   Tokyo 181-0015, Japan.}
\altaffiltext{14}{Institute for Cosmic Ray Research, University
   of Tokyo, Kashiwa, 2778582, Japan.}
\altaffiltext{15}{The Institute for Advanced Study, Princeton,
   NJ 08540.}
\altaffiltext{16}{Microsoft Research, 455 Market Street, Suite 1690,
   San Francisco, CA 94105.}
\altaffiltext{17}{Department of Physics and Astronomy, The Johns
   Hopkins University, 3400 North Charles Street, Baltimore, MD 21218.}
\altaffiltext{18}{Mt. Suhora Observatory, Cracow Pedagogical
     University, ul. Podchorazych 2, 30-084, Cracow, Poland.}
\altaffiltext{19}{Astronomy Centre, University of Sussex, Falmer,
   Brighton BN1~9QJ, UK.}
\altaffiltext{20}{Space Telescope Science Institute,
   3700 San Martin Drive, Baltimore, MD 21218.}
\altaffiltext{21}{US Naval Observatory, Flagstaff Station,
   P.O. Box 1149, Flagstaff, AZ 86002-1149.}
\altaffiltext{22}{Department of Physics, Applied Physics and
   Astronomy, Rensselaer Polytechnic Institute, Troy, NY 12180.}
\altaffiltext{23}{Institute of Cosmology and Gravitation, Mercantile
   House, Hampshire Terrace, University of Portsmouth, Portsmouth, PO1 2EG, UK.}
\altaffiltext{24}{UCO/Lick Observatory, University of California,
     Santa Cruz, CA 96064.}
\altaffiltext{25}{Lawrence Berkeley National Laboratory,
     Physics Division, 1 Cyclotron Road, Berkeley, CA 94720.}
\altaffiltext{26}{Carnegie Observatories, 813 Santa Barbara
Street, Pasadena, CA 91101.}
\altaffiltext{27}{Enrico Fermi Institute, The University of Chicago,
  5640 South Ellis Avenue, Chicago, IL 60637.}


\begin{abstract}
We present the third edition of the Sloan Digital Sky Survey (SDSS)
Quasar Catalog.  The catalog consists of the~46,420 objects
in the SDSS Third Data Release that have luminosities larger
than \hbox{$M_{i} = -22$} (in a cosmology with
\hbox{$H_0$ = 70 km s$^{-1}$ Mpc$^{-1}$,}
\hbox{$\Omega_M$ = 0.3,}
and \hbox{$\Omega_{\Lambda}$ = 0.7),} have at least one
emission line with FWHM larger than 1000~km~s$^{-1}$ 
or are unambiguously broad absorption line quasars, 
are fainter than \hbox{$i = 15.0$,}
and have highly
reliable redshifts.
The area covered by the catalog is~$\approx$~4188~deg$^2$.
The quasar redshifts range from~0.08 to~5.41, with a median value of~1.47;
the high-redshift sample includes 520 quasars at
redshifts greater than four, of which 17 are at redshifts greater than five.
For each object the catalog presents positions accurate
to better than~0.2$''$~rms per coordinate,
five-band ($ugriz$) CCD-based photometry with typical accuracy
of~0.03~mag, and information on the morphology and selection method.
The catalog also contains radio, near-infrared, and X-ray emission
properties of the quasars, when available, from other large-area surveys.
The calibrated digital spectra cover the wavelength region
3800--9200 \AA\ at
a spectral resolution \hbox{of $\simeq$ 2000;} the spectra can be retrieved
from the public database using the information provided in the catalog.
A total of~44,221 objects in the catalog were discovered by the SDSS;
28,400 of the SDSS discoveries are reported here for the first time.

\end{abstract}

\keywords{catalogs, surveys, quasars:general}

\section{Introduction}

This paper describes the Third Edition of the Sloan Digital Sky Survey
(SDSS; York et al.~2000) Quasar Catalog.  The first two editions,
coinciding with the SDSS Early Data Release (EDR; Stoughton et al.~2002) and
the SDSS First Data Release (DR1; Abazajian et al.~2003), contained 3,814 and
16,713 quasars, respectively (Schneider et al.~2002, 2003; hereafter
Papers~I and~II).  The current catalog
contains 46,420 quasars from the SDSS Third Data Release
(DR3; Abazajian et al.~2005), of which 28,400 (61\%)
are presented here for the first time.  The number of new quasars reported
here is comparable to the number in the final 2dF QSO Redshift Survey
(2QZ) catalog
(Croom et al.~2004); the total size is similar to that of the
NASA/IPAC Extragalactic Database (NED) Quasar Catalog.

The catalog in the present paper consists of the DR3 objects that
have a luminosity larger than
\hbox{$M_{i} = -22.0$}  (calculated assuming an
\hbox{$H_0$ = 70 km s$^{-1}$ Mpc$^{-1}$,} \hbox{$\Omega_M$ = 0.3,} 
\hbox{$\Omega_{\Lambda}$ = 0.7} cosmology [Spergel et al.~2003],
which will be used throughout this paper), and whose SDSS
spectra contain at least one broad emission line
(velocity FWHM larger than \hbox{$\approx$ 1000 km s$^{-1}$)}
or are unambiguously broad absorption line quasars.
The catalog also has a bright limit \hbox{of $i = 15.0$.}
The quasars range in redshift from~0.08 to~5.41, and
44,221~(95\%) were discovered by the SDSS.

The objects are denoted in the catalog by their DR3
J2000 coordinates;
the format for the object name
is \hbox{SDSS Jhhmmss.ss+ddmmss.s}.  Since continual improvements
are being made to the SDSS data processing software, the astrometric
solutions to a given set of observations can result in modifications to the
coordinates of an object at the~0.1$''$ to~0.2$''$ level, hence
the designation of a given source can change between data releases.  Except on
very rare occasions (see \S 5.1), this change in position is much less
than~1$''$.
When merging SDSS Quasar Catalogs with previous databases one should
always use the coordinates, not object names, to identify unique entries.


The DR3 catalog does not include classes of Active Galactic Nuclei (AGN)
such as Type~II quasars, Seyfert galaxies, and BL~Lacertae objects; studies
of these sources
in the SDSS
can be found in Zakamska et al.~(2003) (Type II), Kauffmann et al.~(2003)
and Hao et al.~(2005) (Seyferts), and Anderson et al.~(2003)
and Collinge et al.~(2005) (BL Lacs).  Spectra of the
highest redshift SDSS quasars \hbox{($z > 5.7$;}
e.g., Fan et al.~2003) were not acquired as
part of the SDSS survey, so they are not included in the catalog.

The observations used to produce the catalog are presented in
\S 2; the construction of the catalog and the catalog format
are discussed in \S\S 3 and~4, respectively.  Section~5
contains an overview of the catalog, and a brief description of future work is
given in \S 6.
The catalog is presented in an electronic table in this paper and
can also be found at an SDSS public web
site.\footnote{\tt 
http://www.sdss.org/dr3/products/value$\_$added/qsocat$\_$dr3.html}

\section{Observations}

\subsection{Sloan Digital Sky Survey}

The Sloan Digital Sky Survey
uses a CCD camera \hbox{(Gunn et al. 1998)} on a
dedicated 2.5-m telescope
at Apache Point Observatory,
New Mexico, to obtain images in five broad optical bands ($ugriz$;
Fukugita et al.~1996) over approximately
10,000~deg$^2$ of the high Galactic latitude sky.
The
survey data processing software measures the properties of each detected object
in the imaging data in all five bands, and determines and applies both
astrometric and photometric
calibrations (Pier et al., 2003; \hbox{Lupton et al. 2001};
Ivezi\'c et al.~2004).
Photometric calibration is provided by simultaneous
observations with a 20-inch telescope at the same site (see Hogg
et al.~2001, Smith et al.~2002, and Stoughton et al.~2002).
The SDSS photometric system is based on the AB magnitude scale
(Oke \& Gunn~1983).

The catalog contains photometry from~136 different
SDSS imaging runs acquired between 
1998~September~19 (Run~94) and 2003~May~1 (Run~3927)
and spectra from 826 spectroscopic plates taken between 
2000~March~5 and 2003~July~6.

\subsection{Target Selection}

The SDSS filter system was designed to identify quasars at redshifts between
zero and approximately six (see Richards et al.~2002);
most quasar candidates are selected based on
their location in multidimensional SDSS color-space.
The Point Spread Function (PSF) magnitudes are used for the quasar
target selection, and the selection is based on magnitudes and colors
that have been corrected for Galactic extinction
(using the maps of Schlegel, Finkbeiner, \& Davis~1998).
An $i$ magnitude limit of~19.1
is imposed for candidates whose colors indicate
a probable redshift of less than~$\approx$~3 (selected from the $ugri$
color cube);
high-redshift candidates (selected from the $griz$ color cube)
are accepted if \hbox{$i < 20.2$.}  The errors on the $i$ measurements
are typically \hbox{0.02--0.03} and \hbox{0.03--0.04} magnitudes at the
brighter and fainter limits, respectively.
The SDSS images of the high-redshift candidates must be unresolved.
In addition to the multicolor selection, unresolved objects brighter
\hbox{than $i = 19.1$} that lie within~2.0$''$ of a FIRST radio source
(Becker, White, \&~Helfand~1995) are also identified as primary quasar
candidates.
A detailed description of the quasar selection process and possible
biases can be
found in Richards et al.~(2002) and Paper~II.

Supplementing the primary quasar sample described above are
quasars that were targeted by
the following SDSS software selection packages:
Galaxy (Strauss et al.~2002 and
Eisenstein et al.~2001),
X-ray (object near the position of a {\it ROSAT} All-Sky Survey
[RASS; Voges et al.~1999,~2000]
source; see Anderson et al.~2003),
Star (point source with unusual color), or Serendipity (unusual color
or FIRST matches).
No attempt at completeness was made for the last three categories.
Most of the DR3 quasars that
fall below the magnitude limits of the quasar survey were selected by
the serendipity algorithm (see \S 5).

Target selection also imposes a maximum brightness limit
\hbox{($i = 15.0$)} on quasar candidates; the spectra of objects that
exceed this brightness would contaminate the adjacent spectra 
on the detector of the SDSS spectrographs.

One of the most important tasks during the SDSS commissioning period
was to refine the quasar target selection algorithm (see Papers~I and~II);
some of the DR3 data (and all of the material in Paper~II)
were taken before the quasar selection algorithm as described in
Richards et al.~(2002) was implemented.
Once the final target selection software was
installed, the algorithm was applied to the entire SDSS photometric database,
and the DR3 quasar catalog lists the selection target flag for each object
produced by the final selection algorithm. 
Most of the quasars that have been added to the catalog since the DR1 version
were found with the Richards et al.~(2002) algorithm.

It is important to note
that extreme care must be exercised when constructing statistical samples
from this catalog; if one uses the values produced by only the latest version
of the selection software, not only must one drop known quasars that were not
identified as quasar candidates by the final selection software, one must also
account for
quasar candidates produced by the final version that were not observed in the
SDSS spectroscopic survey (this can occur
in regions of sky whose spectroscopic targets were
identified by early versions of the selection software).
The selection for the UV-excess quasars,
which comprise the majority ($\approx$ 80\%) of the objects in the
DR3 Catalog, has remained
reasonably uniform; the changes to the selection algorithm were primarily
designed to increase the effectiveness of the identification of
\hbox{$3.0 < z < 3.8$} quasars.
Extensive discussions of the
completeness and efficiency of the selection can be found in
Vanden~Berk et al.~(2005) and
Richards et al.~(2005); the latter paper discusses the issues that are
important for the construction of 
statistical SDSS quasar samples.  The survey efficiency (the ratio of
quasars to quasar candidates) for the $ugri$-selected candidates, which comprise
the bulk of the quasar sample, is about~75\%.

\subsection{Spectroscopy}

Spectroscopic targets chosen by the various SDSS selection algorithms
(i.e., quasars, galaxies, stars, serendipity) are arranged onto
a series of 3$^{\circ}$ diameter circular fields (Blanton et al.~2003).
Details of the spectroscopic observations can be found in
York et al.~(2000), Castander et al.~(2001), Stoughton et al.~(2002),
and Paper~I.
There are~826 DR3 spectroscopic fields;
the locations of the plate centers
can be found from the information given by Abazajian et al.~(2005).
The DR3 spectroscopic program attempted to cover, in a well defined manner,
an area of~$\approx$~4188~deg$^2$.  Spectroscopic plate~716 was the first
spectroscopic observation that was based on the
final version of the quasar target selection algorithm (Richards et al.~2002);
however, the detailed tiling
information in the SDSS database must be consulted to identify those regions
of sky targeted with the final selection algorithm.

The two double-spectrographs produce data covering \hbox{3800--9200 \AA }
at a spectral resolution \hbox{of $\simeq$ 2000.}
The data, along with the associated calibration frames, are processed by
the SDSS spectroscopic pipeline (see Stoughton et al.~2002).
The calibrated spectra are classified into various groups
(e.g., star, galaxy, quasar), and redshifts are determined by two independent
software packages.
Objects whose spectra cannot be classified
by the software are flagged for visual inspection.
Figure~1 shows the calibrated SDSS spectra of four previously unknown
catalog quasars representing a range of properties.

The spectrophotometric calibration has been considerably improved since
DR1; details of the changes are given in Abazajian et al.~(2004).
The processed DR3 spectra {\it have not} been corrected for Galactic
extinction; this is a change from all previous SDSS data releases.

\section{Construction of the SDSS Quasar Catalog}

The quasar catalog was constructed in three stages: 1)~Creation of a
quasar candidate database, 2)~Visual examination of the candidates'
spectra, and 3)~Application of luminosity
and emission-line velocity width criteria.

Because of the evolution of the project software during the early phases
of the SDSS, spectra of quasars could be obtained based on photometric
measurements/target selection criteria that
are not identical to the final products.
The DR3 catalog was prepared using Version~5.4 of the photometric pipeline
code.
(The differences between this release of the photometric software and previous
versions are described in Abazajian et al.~2004.)
These photometric measurements are referred to as BEST values;
the measurements used during the target selection
are denoted as TARGET values.  Differences between TARGET and BEST arise when
1)~the image data are processed with a new version of the software or 2)~new
image data replace the observations used for the target selection.
See Paper~II for a more extensive
discussion of the differences between TARGET and BEST and the impact this
has on the SDSS quasar catalogs.

The absolute magnitude limit (\S~3.3) was imposed on 
the BEST photometry.  However, we also report the TARGET photometry, which 
may be more useful when constructing statistical samples.
This situation arises because 
small changes in the photometry, while leaving the {\em density} of 
quasars constant, can change the {\em individual} quasars that appear in 
the sample; thus only the TARGET sample has sufficient spectroscopic 
completeness in terms of statistical analysis.

\subsection{Creation of the Quasar Candidate Database}

The construction of the DR3 Quasar Catalog began with four separate queries
(three automated, one manual) to the SDSS Catalog Archive
Server\footnote{\tt http://cas.sdss.org} (CAS).  These queries produced
a set of objects that should contain all of the quasars found by the SDSS;
various automated and interactive culls were applied to this sample
to create the final catalog.

Before describing the queries in detail, we must examine the definition
of the term ``primary object'' within the SDSS object catalog, as it
is relevant to some of the queries.  The CAS defines a
unique set of object detections in order to remove duplications
(e.g., an object can be detected twice in the overlap area of neighboring
runs).  This unique set of objects is designated as ``primary'' in
the database, and only ``primary'' objects are considered during
target selection (the remaining objects can be ``secondary'' or
``family''; see \S4.7 in Stoughton et al.~(2002) for a definition of these
terms).  Due to differences in the photometric pipeline between the
TARGET and BEST mentioned in the previous section, it is possible that
the BEST object
belonging to an existing spectrum is not designated ``primary'', or is
missing from the BEST catalog altogether.  This can occur when
different data are used for TARGET and BEST, or because of changes in the
photometric pipeline, in particular the deblender (see \S 4.4 of
Abazajian et al.~2004). In addition, a few plates cover sky
area outside the
nominal survey boundaries (the so-called ``bonus plates''\footnote{
\tt http://www.sdss.org/dr3/products/spectra/}); all objects on these
plates are non-primary in the TARGET version, and objects that fall outside
of the final survey boundaries remain non-primary in BEST.  The spectroscopic
target selection flags and photometry for both BEST and TARGET processing
are included in the catalog; the latter set are important for constructing
statistical samples (see Richards et al.~2005).

Below we present a brief description of the four queries used to assemble
the quasar candidate database.  The actual text of the queries is given
following each description.  Since some minor changes to the DR3 database
have been made since the the start of the construction of the catalog
(e.g., manually revised redshifts and spectral classifications),
queries run on the
latest DR3 database may return slightly different numbers of objects than
quoted below.

\smallskip\noindent
1. The first query is the union of 
primary objects targeted as quasar candidates and 
primary objects not targeted as quasar candidates but
whose spectra were either classified by the SDSS spectroscopic software
as {\tt QUASAR}, had
redshifts~$\ge$~0.6, or were
unidentified by the automated software (SDSS spectral class {\tt UNKNOWN}).
This query produced~130,119 objects; this is the
vast majority (over 99.5\%) of the initial quasar database.

\medskip

\vbox{
\noindent{\tt SELECT * FROM BESTDR3..photoObjAll as p\\
left outer join SpecObjAll as s on p.objID = s.bestObjID\\
WHERE ( (p.mode = 1) AND ( (p.primTarget \& 0x0000001f) $>$ 0 OR\\
( (p.primTarget \& 0x0000001f) = 0 AND (s.z$>$=0.6 OR
s.specClass in (3,4,0)) ) ) )}
}

\smallskip\noindent
2. The second query recovered~286 objects with quasar or {\tt UNKNOWN}
spectra that were
mapped to a photometric object in TARGET but not in BEST (due to differences in 
deblending, etc., between the two pipelines).

\medskip

\vbox{
\noindent{\tt SELECT * FROM BESTDR3..SpecObjAll as s\\
join TARGDR3..PhotoObjAll as p on p.objID = s.targetObjID\\
WHERE (s.bestObjID=0 AND (zWarning \& 0x00020000 = 0) AND (s.specClass in
(3,4,0)))}
}

\smallskip\noindent
3. Our third query was designed to recover non-primary objects with spectral
classification of {\tt QUASAR} or {\tt UNKNOWN}.  There are~251 such objects
in the quasar database, mostly from the ``bonus plates".

\medskip

\vbox{
\noindent{\tt SELECT * FROM BESTDR3..specObjAll as s\\
left outer join photoObjAll as p on s.bestObjID = p.objID\\
WHERE ( (p.mode in (2,3)) AND (s.specClass in (3,4,0)) )}
}

\smallskip\noindent
4. The previous three automated queries missed 33 quasars from the DR1 catalog.
Four of these objects were database glitches in DR1 - the objects are
definitely quasars, but it is not possible to map properly these spectra
to the spectroscopic fibers, so we cannot be certain of these
quasars' celestial coordinates.  We were able to identify positively one
of these four quasars, as it was discovered by the Large Bright Quasar
Survey (Hewett, Foltz, and Chaffee 1995), but the other three are lost.
The remaining~29 objects were not targeted as quasars in BEST and
the software processing incorrectly
did not classify the spectrum as {\tt QUASAR} or
{\tt UNKNOWN}; these objects were added into the quasar candidate database.
The existence of these quasars suggests an $\simeq$0.2\%
incompleteness rate in our cataloging of quasars in post-DR1 data.  Since
this query simply retrieved information on specific DR1 catalog quasars not
recovered by the above three queries, the text of this query is not given
here.

In an ideal survey (e.g., one where there was no repeat imaging,
no area overlaps, no change in software) only the first query
would be required.

Three automated cuts were made to the raw quasar database of approximately
131,000 candidates:
1)~the over 58,000 objects targeted as quasars but whose
spectra had not yet been obtained by the closing date of DR3
were deleted, 2)~candidates
classified with high confidence as ``stars" and had redshifts less
than~0.002 were rejected
(7647 objects),\footnote{After this paper was submitted, we realized we
could estimate the number of star+quasar blends discarded in this step using
FIRST counterparts.  Two radio-detected star+quasar blends missing from the
catalog were discovered (SDSS J092853.5+570735.3 at \hbox{$z = 1.67$}
and SDSS J105115.8+464417.3 at \hbox{$z = 1.42$}).
Given that only 8\% of catalog objects are detected by FIRST, we expect that
$25^{+33}_{-16}$ additional FIRST-undetected quasars are missing from the
catalog due to blending with stars.  Overall, star+quasar blending appears
to create a negligible incompleteness of 0.06$^{+0.07}_{-0.03}$\%.}
and 3)~multiple observations of the same object
(coordinate agreement better than~1$''$) were resolved;
the primary spectrum with the
highest S/N ratio was retained (this action deleted 1205 spectra).
These culls produced a list of~63,614
unique quasar candidates.

\subsection{Visual Examination of the Spectra}

The SDSS spectra of the remaining quasar
candidates were manually inspected by
several of the authors (DPS, PBH, GTR, MAS, DVB, and SFA).
This effort confirmed that the
spectroscopic pipeline redshifts and classifications
of the overwhelming majority of the objects
are accurate.
Several thousand objects were dropped from the
list because they were obviously not quasars (these objects tended to be
low S/N stars, unusual stars, and a mix of absorption-line and
narrow emission-line galaxies).
Spectra for which redshifts could not be determined (low signal-to-noise
ratio or subject to data-processing difficulties) were also removed from
the sample.
This visual inspection resulted in the revisions of
the redshifts of a few hundred quasars; this change in the redshift was usually
quite substantial.  Most of these corrections have been applied to the CAS.

About~1\% of the entries in the catalog (a few hundred objects)
are not ``ironclad" classical quasars or lack absolutely certain redshifts.
There are numerous ``extreme Broad Absorption Line (BAL) Quasars"
(see Hall et al.~2002, 2004); it is difficult
if not impossible to apply the emission-line width criterion for these objects,
but they are clearly of interest, have more in common with ``typical" quasars
than with narrow-emission line galaxies, 
and have historically been included in quasar
catalogs.  We have included in the catalog all objects with broad
absorption-line spectra that meet the \hbox{$M_i < -22$} luminosity criterion.
The spectra at the S/N limit of the catalog occasionally yield likely
but not certain redshifts (witness the revisions of the redshifts of
a few objects in each edition of this series of papers; see \S 5.1).

\subsection{Luminosity and Line Width Criteria}

As in Paper~II, we adopt a luminosity limit of
\hbox{$M_{i} = -22.0$} for an
\hbox{$H_0$ = 70 km s$^{-1}$ Mpc$^{-1}$,} \hbox{$\Omega_M$ = 0.3,} 
\hbox{$\Omega_{\Lambda}$ = 0.7} cosmology (Spergel et al.~2003).
The absolute magnitudes were calculated by correcting the $i$ 
measurement for Galactic extinction (using the maps of
Schlegel, Finkbeiner, \& Davis~1998) and assuming that the quasar
spectral energy distribution in the ultraviolet-optical
can be represented by a power law
\hbox{($f_{\nu} \propto \nu^{\alpha}$),} where $\alpha$~=~$-0.5$
(Vanden~Berk et al.~2001).  This calculation ignores the contributions
of emission lines and the observed distribution in continuum slopes.
Emission lines can contribute several tenths of a magnitude to the
k-correction (see Richards et al 2001), and variations in the continuum
slopes can introduce a magnitude or more of error into the calculation
of the absolute magnitude, depending upon the
redshift.
The absolute magnitudes
will be particularly uncertain at redshifts near and above
five when the Lyman~$\alpha$ line (with a typical observed equivalent width
of~$\approx$~500~\AA ) and strong Lyman~$\alpha$ forest absorption enter
the~$i$ bandpass.

Our catalog has a luminosity limit of~$M_i$~=~$-22.0$, which is
lower than the cutoff in most quasar catalogs (see Paper~II for a discussion
of this point).
Objects near this limit can have an appreciable amount of contamination
by starlight (the host galaxy).  Although the SDSS photometric measurements
in the catalog are based on the PSF magnitudes, the nucleus of the host
galaxy can appreciably contribute to this measurement for the lowest luminosity
entries in the catalog (see Hao et al.~2005).
An object of $M_i = -22.0$ will reach the \hbox{$i = 19.1$} ``low-redshift"
selection limit at a redshift of~$\approx$~0.4.

After visual inspection and application of the luminosity criterion had
reduced the number of quasar candidates to under 50,000 objects, the
remaining spectra were processed with an automated line measuring routine.  The
spectra for objects whose maximum line width was less than 1000~km~s$^{-1}$
were visually examined; if the measurement was deemed to be an accurate
reflection of the line (automated routines occasionally have spectacular
failures when dealing with complex line profiles), the object was removed
from the catalog.  The resulting catalog contains~46,420 entries.

\section{Catalog Format}

The DR3 SDSS Quasar Catalog is available in three types of files at an
SDSS public web site:
1)~a standard ASCII file with fixed-size columns,
2)~a gzipped compressed version of the ASCII file (which is smaller than
the uncompressed version
by a factor of nearly five), and 3)~a binary FITS table format.
The following description applies to the standard ASCII file.  All files
contain the same number of columns, but the storage of the numbers differs
slightly in the ASCII and FITS formats; the FITS header contains all of the
required documentation.  Table~1 provides a summary of the information
contained in each of the columns in the catalog.

The standard ASCII catalog (Table~2 of this paper)
contains information on~46,420 quasars in
a 19.8~megabyte file.
The DR3 format is similar to that of DR1; the major difference is
the inclusion
of some additional SDSS observational/processing material in the DR3
catalog.

The first~71 lines consist of catalog documentation; this is followed
by~46,420 lines containing
information on the quasars.  There are~65 columns in each line; a summary
of the information is given in Table~1 (the documentation in the ASCII catalog
header
is essentially an expansion of Table~1).  At least one space separates all the
column entries, and, except for the first and last columns (SDSS and NED
object names), all entries are reported in either floating point or
integer format.

Notes on the catalog columns:

\noindent
1) The DR3 object designation, given by the format
\hbox{SDSS Jhhmmss.ss+ddmmss.s}; only the final~18
characters (i.e., the \hbox{``SDSS J"} for each entry is dropped)
are listed in the catalog.

\noindent
2--3) The J2000 coordinates (Right Ascension and
Declination) in decimal degrees.  The positions for the vast majority of
the objects are accurate to~0.1$''$~rms or better
in each coordinate; the largest
expected errors are~0.2$''$ (see Pier et al~2003).  The SDSS coordinates
are placed in the International Celestial Reference System, primarily
through the USNO CCD Astrograph Catalog (Zacharias et al.~2000), and
have an rms accuracy of~0.045$''$ per coordinate.

\noindent
4) The quasar redshifts.
A total of 377 of the CAS redshifts were revised during our visual inspection.  
A detailed description of the redshift measurements is given in Section 4.10
of Stoughton et al.~(2002).  A comparison of 299 quasars observed
at multiple epochs by the SDSS (Wilhite et al.~2005) finds an rms
difference of~0.006 in the measured redshifts for a given object.
%

\noindent
5--14) The DR3 PSF
magnitudes and errors (not corrected for Galactic reddening) from BEST
photometry (or, when BEST is unavailable, from TARGET photometry)
for each object in the five SDSS filters.
The effective wavelengths of the $u$, $g$, $r$, $i$, and $z$ bandpasses
are 3541, 4653, 6147, 7461, and~8904~\AA, respectively (for
\hbox{$\alpha = -0.5$} power-law spectral energy distribution using the
definition of effective wavelength given in Schneider, Gunn, and
Hoessel~1983).
The photometric measurements are reported
in the natural system of the SDSS camera, and
the magnitudes are normalized to the
AB system (Oke \& Gunn~1983).
The measurements are reported as
asinh magnitudes (Lupton, Gunn, \& Szalay~1999); see Paper~II and
Abazajian et al.~(2004) for additional
discussion and references for the accuracy of the photometric measurements.
The TARGET photometric measurements are presented in columns \hbox{55--64.}

\smallskip\smallskip
\vbox{\noindent
15) The Galactic extinction in the $u$ band based on the maps of
Schlegel, Finkbeiner, \& Davis~(1998).  For an $R_V = 3.1$ absorbing medium,
the extinctions in the SDSS bands can be expressed as

$$ A_x \ = \ C_x \ E(B-V) $$

\noindent
where $x$ is the filter ($ugriz$), and values of $C_x$ are
5.155, 3.793, 2.751, 2.086, and 1.479 for $ugriz$, respectively
($A_g$, $A_r$, $A_i$, and $A_z$ are 0.736, 0.534, 0.405, and 0.287 times
$A_u$).
}

\noindent
16) The logarithm of the Galactic neutral hydrogen column density along the
line of sight to the quasar. These values were
estimated via interpolation of the 21-cm data from Stark et al.~(1992),
using the COLDEN software provided by the {\it Chandra} X-ray Center.
Errors associated with the interpolation are typically expected to
be less than $\approx 1\times 10^{20}$~cm$^{-2}$ (see \S5 of
Elvis, Lockman, \& Fassnacht 1994).

\noindent
17) Radio Properties.  If there is a source
in the FIRST catalog
within~2.0$''$ of
the quasar position, this column contains the FIRST
peak flux density at 20~cm encoded as an AB magnitude

$$ AB \ = \ -2.5 \log \left( {f_{\nu} \over 3631 \ {\rm Jy}} \right) $$

\noindent
(see Ivezi\'c et al.~2002).
An entry of ``0.000" indicates no match to a FIRST source; an entry of
``$-1.000$" indicates that the object does not lie in the region covered by
the final catalog of the FIRST survey.

\noindent
18) The S/N of the FIRST source whose flux is given in column~17.

\noindent
19) Separation between the SDSS and FIRST coordinates (in arc seconds).

\noindent
20-21)
These two columns provide information about
extended FIRST counterparts to SDSS quasar so as to identify
some of the potentially most interesting extended radio
sources in the catalog.

In cases when the FIRST counterpart to
an SDSS source is extended, the FIRST catalog position of the source
may differ by more than 2$''$ from the optical position. A~``1" in column~20
indicates that no matching FIRST source was found within
2$''$ of the optical position, but that there {\it is}
significant detection (larger than~3$\sigma$)
of FIRST flux at the optical position. This is
the case for 1319 SDSS quasars.

A ``1" in column~21 identifies
the 891 sources with a FIRST match in either column~17 or~20 that also
have at least one FIRST counterpart located between 2.0$''$ (the SDSS-FIRST
matching radius) and
30$''$ of the optical position.
Based on the average FIRST
source surface density of 90~deg$^{-2}$, we \hbox{expect 20--30} of these
matches to be chance superpositions.

\noindent
22) The logarithm
of the vignetting-corrected count rate (photons s$^{-1}$)
in the broad energy band \hbox{(0.1--2.4 keV)} in the
{\it ROSAT} All-Sky Survey Faint Source Catalog (Voges et al.~2000) and the
{\it ROSAT} All-Sky Survey Bright Source Catalog (Voges et al.~1999).
The matching radius was set to~30$''$;
an entry of~``$-9.000$" in this column indicates no X-ray detection.

\noindent
23) The S/N of the {\it ROSAT} measurement.

\noindent
24) Separation between the SDSS and {\it ROSAT} All-Sky Survey
coordinates (in arc seconds).

\noindent
25--30) The $JHK$ magnitudes and errors from the
2MASS All-Sky Data Release Point Source Catalog (Cutri et al.~2003) using
a matching radius
of~2.0$''$.  A non-detection by 2MASS is indicated by a ``0.000" in these
columns.  Note that the 2MASS measurements are Vega-based, not AB,
magnitudes.

\noindent
31) Separation between the SDSS and 2MASS coordinates (in arc seconds).

\noindent
32) The absolute magnitude in the $i$ band calculated by correcting for
Galactic extinction and assuming
\hbox{$H_0$ = 70 km s$^{-1}$ Mpc$^{-1}$,}
$\Omega_M$~=~0.3, $\Omega_{\Lambda}$~=~0.7, and a power-law (frequency)
continuum index of~$-0.5$.

\noindent
33) Morphological information.
If the SDSS photometric pipeline classified the image of the quasar
as a point source, the catalog entry is~0; if the quasar is extended, the
catalog entry is~1.

\noindent
34) The SDSS {\tt SCIENCEPRIMARY} flag, which
indicates whether the spectrum was taken as a normal science spectrum
({\tt SCIENCEPRIMARY}~=~1) or for another purpose
({\tt SCIENCEPRIMARY}~=~0).  The latter category contains
Quality Assurance and calibration spectra, or spectra of objects
located outside of the nominal survey area
(e.g., ``bonus" spectra, see \S 3.1).

\noindent
35) The SDSS {\tt MODE} flag, which provides information on
whether the object is designated primary ({\tt MODE} = 1), secondary
({\tt MODE} = 2), or family ({\tt MODE} = 3).  During target selection,
only objects with {\tt MODE} = 1
are considered (except for objects on ``bonus" plates);
however, differences between TARGET and BEST
photometric pipeline versions make it possible that the BEST
photometric object belonging to a spectrum is either not detected at
all, or is a non-primary object (see \S 3.1 above).  Over~99.5\% of the
catalog entries are primary;
174 quasars are secondary and~6 are family.
For statistical analysis, users should restrict 
themselves to primary objects; secondary and 
family objects are included in the catalog for the sake of completeness
with respect to confirmed quasars.

\noindent
36) The 32-bit SDSS target selection flag from BEST processing
({\tt PRIMTARGET}; see Table~26 in Stoughton et al. 2002 for details).
The target selection flag from TARGET processing is found in column~54.

\noindent
37-43) The spectroscopic target selection status (BEST) for each object.
The target selection flag in column~36 is decoded for seven groups:
Low-redshift quasar, High-redshift quasar, FIRST, ROSAT, Serendipity,
Star, and Galaxy; see Table~3 for a summary.
An entry of~``1" indicates that the object satisfied the given criterion
(see Stoughton et al.~2002).  Note that an object can
be targeted by more than one selection algorithm.

\noindent
44--45) The SDSS Imaging Run number and the Modified Julian Date (MJD) of the
photometric observation used in the catalog.  The MJD is given as an integer;
all observations on a given night have the same integer MJD
(and, because of the observatory's location, the same UT date). For example,
imaging run 94 has an MJD of 51075; this observation was taken on the 
night of 1998 September~19~(UT).

\noindent
46--48) Information about the spectroscopic observation (Modified Julian
Date, spectroscopic plate number, and spectroscopic fiber number) used to
determine the redshift.
These three numbers are unique for each spectrum, and
can be used to retrieve the digital spectra from the public SDSS database.

\noindent
49--53) Additional SDSS processing information: the
photometric processing rerun number; the camera column (1--6) containing
the image of the object, the frame number of the run containing the object,
the object identification number, and the ``chunk" number (referred to as
`tilerun' in the CAS) used to assign
the target selection flag (see Stoughton et al.~2002 for descriptions of
these parameters).

\noindent
54) The 32-bit SDSS target selection flag from the TARGET processing.

\noindent
55--64) The DR3 PSF
magnitudes and errors (not corrected for Galactic reddening) from TARGET
photometry.  For 59 quasars, the $u$ TARGET information is missing from 
the CAS due to a software error;
these objects have "0.000" entered for the $u$~TARGET values.

\noindent
65) NED information.
If there is a source in the NED quasar database within~5.0$''$ of the
quasar position, the NED object name is given in this column.
The matching was done using
the 45,526 objects in the NED quasar database as of August~2004.

\section{Catalog Summary}

Of the 46,420 objects in the catalog, 44,221 were discovered by the SDSS, and
28,400 are presented here for the first time.
(We classify an object as
previously known if the NED Quasar Catalog contains a quasar
within~5$''$ of the SDSS position.
Occasionally NED lists the SDSS designation for an object that was discovered
earlier via another investigation; we have not attempted
to correct these misattributions.)
The catalog quasars span a wide range of properties: redshifts
from~0.078 to~5.414, \hbox{$ 15.10 < i < 21.78$}
(160~objects \hbox{have $i > 20.5$;} only five
have \hbox{$i > 21.0$}),
and \hbox{$ -30.2 < M_{i} < -22.0$.}
The catalog contains 3761, 2672, and~6192
matches to the FIRST, RASS, and 2MASS catalogs, respectively.
The RASS and 2MASS catalogs cover essentially all of the DR1 area, but~4683
(10\%) of
the entries in the DR3 catalog lie outside of the FIRST region.

Figure~2 displays the distribution of the DR3 quasars in the $i$-redshift plane
(the five objects with \hbox{$i > 21$} are not plotted).
Objects which NED indicates were previously discovered by investigations other
than the SDSS
are indicated with open circles.  The curved cutoff on the left
hand side of the graph is due to the minimum luminosity criterion
\hbox{($M_i < -22$).}  The ridge in the contours at
\hbox{$i \approx 19.1$}
for redshifts below three reflects the flux limit of the
low-redshift sample; essentially all of the
large number of \hbox{$z < 3$} points with \hbox{$i > 19.1$}
are quasars selected via criteria other than the primary
multicolor sample.

A histogram of the catalog redshifts is shown in Figure~3.  A clear
majority of
quasars have redshifts below two (the median redshift is~1.47, the
mode is~$\approx$~1.85),
but there is a significant tail
of objects out beyond a redshift of five
\hbox{($z_{\rm max}$ = 5.41).}  The dips in the curve at redshifts
of~2.7 and~3.5 arise because the SDSS colors of quasars at these redshifts
are similar to the colors of stars; we decided to accept significant
incompleteness at these redshifts rather than be overwhelmed by a large number
of stellar contaminants in the spectroscopic survey.  Improvements in the
quasar target selection algorithm since the previous edition of the
SDSS Quasar Catalog have considerably increased the efficiency of target
selection at redshifts near~3.5 (compare Figure~3 with Paper~II's
Figure~4; see Richards et al.~(2002) for a discussion).

The distribution of the observed $i$ magnitude
(not corrected for Galactic extinction) of the quasars is given in Figure~4.
The sharp drops in the histogram at \hbox{$i \approx 19.1$} and
\hbox{$i \approx 20.2$} are due to the magnitude limits in the low and
high redshift samples, respectively.

Figure~5 displays the distribution of the absolute~$i$ magnitudes.  There
is a roughly symmetric peak centered at \hbox{$M_i = -26$} with a FWHM
of approximately one magnitude.  The histogram drops off sharply at
high luminosities (only~1.6\% of the objects have \hbox{$M_i < -28.0$)}
and has a gradual decline towards lower luminosities.

A summary of the spectroscopic selection is given in Table~3.  We report
seven selection classes in the catalog (columns~37 to~43).
The second column in Table~3 gives the number of objects that
satisfied a given selection criterion; the third column
contains the number of objects that were identified only by that selection
class.  Slightly over two-thirds~(68\%) of the catalog entries
were selected based on the SDSS
quasar selection criteria (either a low-redshift or high-redshift candidate,
or both).
Approximately~55\% of the quasars
in the catalog are serendipity-flagged candidates,
which is also primarily an ``unusual
color" algorithm; about one-fifth of the catalog was selected by
the serendipity criteria alone.

Of the~31,403 DR3 quasars that have Galactic-absorption corrected
$i$ magnitudes brighter than~19.1,
29,345 (93.4\%) were found from the quasar multicolor
selection; if one combines multicolor and FIRST selection (the primary
quasar target selection criteria),
all but~1777 of the \hbox{$i < 19.1$} objects were selected.

\subsection{Differences Between the DR1 and DR3 SDSS Quasar Catalogs}

The DR1 Catalog (Paper~II) contains~16,713 objects.  The DR3 coverage includes
all of the Paper~II area, so one would expect that all of the Paper~II
quasars would be included in the new edition.  A comparison of the
catalogs, defining a match as a positional coincidence of better than~1$''$,
reveals that~43 Paper~II quasars (0.26\%) are missing in the new catalog.
Each of these cases has been investigated; a summary of the results is
given in Table~4.  There are several reasons for
the omissions:

\noindent
1. Visual examination of the DR3-processed spectrum either convinced us that
the object was not a quasar or that the S/N was insufficient to assign
a redshift with confidence (15 DR1 quasars).

\noindent
2. The widest line in the DR3-processed spectrum had a FWHM of less than
1000~km~s$^{-1}$ (14 DR1 quasars).

\noindent
3. The luminosity of the object dropped below $M_i = -22$.  This can arise
because the latest processing produces new photometric measurements, or
because
different imaging data are used between DR1 and DR3 (in addition to measurement
errors, variability can play a role).  All of the objects dropped for
this reason were near the luminosity cutoff in the DR1 catalog (9 DR1 quasars).

\noindent
4. Uncertain fiber mapping in the DR3 database forced us to drop three
DR1 quasars.  These objects are definitely quasars, but we are no longer
certain (as we thought we were when using the DR1 database)
of the celestial positions.

\noindent
5.  The positions of two DR1 quasars,
whose spectra were taken on spectroscopic plate 540, changed
by more than one arcsecond (in these cases, 2.0$''$ and~3.5$''$) in the
DR3 database.

Four of the entries in the DR3 catalog have redshifts that differ by more
than~0.1 from the DR1 values (the changes in redshift are large: 0.52,
0.84, 1.44, and~2.33).  These quasars are reviewed in \S 5.10.
Only seven quasars have $i$ measurements that differ by more than~0.5
magnitudes between DR1 and~DR3.  In all cases the DR3 measurements are
considered more reliable than those presented in previous publications.

\subsection{Bright Quasars}

Although the spectroscopic survey is limited to objects fainter than
\hbox{$i = 15$}, the SDSS continues to discover ``PG-class"
(Schmidt \& Green~1983) objects.  The DR3 catalog contains 56 entries
with \hbox{$i < 16.0$}; seven were previously unknown.  The spectra of
the brightest two discoveries,
\hbox{SDSS J151921.66+590823.7} \hbox{($i = 15.39$}, \hbox{$z = 0.078$)} and
\hbox{SDSS J152156.48+520238.5} \hbox{($i = 15.44$}, \hbox{$z = 2.21$)},
are presented in Figure~1.  Six of the seven new bright quasars have 
redshifts below~0.2 and are in the
low-luminosity tail of the catalog \hbox{($M_i > -24.0$;} see Figure~5);
but the $z$~=~2.21 object is spectacularly luminous (see \S 5.3).
A comparison of the SDSS and PG surveys is presented in Jester et al.~(2005).

\subsection{Luminous Quasars}

There are 68 catalog quasars with \hbox{$M_i < -29.0$}
(3C~273 has \hbox{$M_{i} \approx -26.6$} in our adopted cosmology); 25 are
published here for the first time.  
HS~1700+6416, \hbox{(= SDSS J170100.62+641209.0)} at
\hbox{$M_i = -30.24$} \hbox{and $z = 2.74$,}
is the most luminous quasar in the catalog.
Four objects have \hbox{$M_i < -30.0$},
including \hbox{SDSS J152156.48+520238.5} (see previous section),
which, at \hbox{$M_i = -30.19$,} is the third most luminous
catalog entry.  The spectrum of this object
possesses a number of low equivalent width
emission lines, which is expected from the Baldwin~(1977) effect.  The
image of the quasar is unresolved, so if it is lensed the image separation
must be considerably less than one~arcsecond (see Pindor et al.~2003).
This object is not seen in the FIRST or the
RASS databases.  The latter point might strike the reader as surprising
given the brightness of the object.  We can quantify the relationship
between the optical brightness and the X-ray upper limit via the quantity
$\alpha_{\rm ox}$, the point-to-point
spectral slope between rest-frame 2500~\AA\ and 2~keV.  For this object,
we find, adopting the assumptions in \S2 of Brandt et~al.~(2002),
\hbox{that $\alpha_{\rm ox} \le -1.7$.}
This constraint is only moderately
interesting; given the luminosity-$\alpha_{\rm ox}$ relation (e.g.,
Strateva et al. 2005), this quasar would be expected to
have $\alpha_{\rm ox}$ just below this limit.
The lack of an X-ray detection could also be explained if
the object were a BAL (e.g., Brandt, Laor, \& Wills~2000), but there
is no evidence of any BAL features in the spectrum.

\subsection{Broad Absorption Line Quasars}

The SDSS Quasar Selection Algorithm has proven to be effective
at finding a wide variety of Broad Absorption Line (BAL) Quasars.
A catalog of 224 BALs drawn from the Paper~I sample is given in
Reichard et al.~(2003); we are currently constructing a BAL catalog, which
will contain well over 1000 objects, from the 46,420 DR3 quasars
(Trump et al. 2005).  BALs are usually recognized by the presence of
C~IV absorption features, which are only visible in SDSS spectra at
$z > 1.6$, thus the frequency of the BAL phenomenon cannot be found from
simply taking the ratio of BALs to total number of quasars in the SDSS
catalog.
During the first few years of the SDSS a wide variety of ``extreme BALs"
were discovered
(see Hall et al.~2002); while the SDSS continues to find significant 
numbers of such objects (the spectrum of a new extreme BAL is displayed
in the lower right panel of Figure~1), the DR3 catalog contains only two
BAL spectra that qualitatively
differ from previous published types: a He~II BAL
\hbox{(SDSSJ162805.81+474415.7)} and a possible BAL with a strange and
unexplained continuum shape
\hbox{(SDSSJ073816.91+314437.1).}  Spectra of both of these BAL quasars
are displayed in Hall et al.~(2004).

\subsection{Quasars with Redshifts Below 0.15}

The catalog contains~69 quasars with redshifts below~0.15;
30 are presented here for the first time.  All of these objects are
of low luminosity \hbox{($M_i > -23.5$)} because of the \hbox{$i = 15.0$}
limit for the spectroscopic sample.  Most of these quasars (53 out of 69)
are extended in the SDSS image data.
Figure~1 displays the spectra of the two lowest redshift quasars among
the recent discoveries,
\hbox{SDSS J151921.66+590823.7} (also mentioned in \S 5.2)
and
\hbox{SDSS J214054.55+002538.2}; both have redshifts near~0.08.

\subsection{High-Redshift ($z \ge 4$) Quasars}

One of the most exciting results produced by the SDSS is
the identification of high-redshift quasars; the SDSS has discovered
quasars out to a redshift of~6.4
(Fan et al.~2003 and references therein).  Quasars with redshifts larger
than~$\approx$~5.7
cannot be found by the SDSS spectroscopic survey because
at these redshifts the observed wavelength of the
Lyman~$\alpha$ emission line is redward of
the $i$ band; at this point quasars become single-filter ($z$) detections.
At the typical $z$-band flux levels for redshift six quasars, there are simply
too many ``false-positives" to undertake automated targeting.
The largest redshift in the DR1 catalog is
\hbox{SDSS J023137.65$-$072854.5} \hbox{at $z = 5.41$}, which was originally
described by Anderson et al.~(2001).  (Indeed, since DR3 represents nearly
half of the survey area, this result suggests that the effective redshift
limit for the SDSS spectroscopic survey is nearer~5.5 than~5.7.)

The DR3 catalog contains 520 quasars with redshifts larger than four; this
is quite striking since but a decade ago the published number of such objects
was only about two dozen.  The SDSS discovered 512 of these quasars;
322 are presented here for the first time.  The
catalog contains~17 quasars with redshifts above five; spectra of the
twelve new objects with the highest redshifts (all with redshifts greater than
or equal to~4.99) are displayed in Figure~6.

The processed spectra for a few of the high-redshift quasars have gaps
(usually caused by extreme contamination of the spectrum from bright,
neighboring objects) that include all of the region containing
the Lyman~$\alpha$ emission line.
The shape of such spectra (in particular the region associated with the
Lyman~$\alpha$ forest), however, are so distinctive
that we are confident that
our redshift assignments are correct.  To verify that this is an appropriate
procedure, we obtained a spectrum with the Low Resolution Spectrograph
(Hill et al.~1998) on the Hobby-Eberly Telescope (HET) of the highest redshift
quasar with this defect in the SDSS spectrum:
\hbox{SDSS~J162623.38+484136.4.}  The HET spectrum confirmed that this
is indeed a redshift~4.9 quasar.

The flux limits of the RASS are such that only the most extreme X-ray
sources can be detected at redshifts larger than four.
We have checked for new X-ray detections in {\it XMM-Newton\/},
{\it ROSAT\/} (pointed observations), and {\it Chandra\/} data;
no additional clear detections are found from the first two instruments,
but six of the $z>4$ quasars in the catalog have previously unreported X-ray
detections in
{\it Chandra\/} data.  These detections have limited numbers of counts, and
thus detailed
X-ray spectral analyses are not possible. We have computed the quasars'
point-to-point
spectral slopes between rest-frame 2500~\AA\ and 2~keV ($\alpha_{\rm ox}$),
adopting
the assumptions in \S2 of Brandt et~al. (2002). Considering the known
dependence of
$\alpha_{\rm ox}$ upon luminosity (e.g., Vignali, Brandt, \& Schneider 2003;
Strateva et~al. 2005), four of the quasars have X-ray emission at a nominal
level
for radio-quiet quasars:
SDSS~J102622.89+471907.0 ($\alpha_{\rm ox}=-1.59$; $z=4.94$),
SDSS~J105322.98+580412.1 ($\alpha_{\rm ox}=-1.57$; $z=5.21$),
SDSS~J222509.19--001406.8 ($\alpha_{\rm ox}=-1.79$; $z=4.89$), and
SDSS~J222845.14--075755.3 ($\alpha_{\rm ox}=-1.78$; $z=5.14$).
These are some of the highest redshift X-ray detections obtained to date.

The remaining two quasars with {\it Chandra\/} detections have more remarkable
X-ray properties.
SDSS~J001115.23+144601.8 ($\alpha_{\rm ox}=-1.28$; $z=4.96$) is a
radio-detected
quasar (37~mJy at 1.4~GHz; Condon et~al. 1998) that is notably X-ray bright.
Its
observed-frame \hbox{0.5--2~keV} flux is
$1.0\times 10^{-13}$~erg~cm$^{-2}$~s$^{-1}$,
making it one of the X-ray brightest objects known at $z>4$. The
basic X-ray and
radio properties of this quasar are similar to those of the handful of X-ray
luminous ``blazars'' studied at $z>4$ (see Table~3 of Bassett et~al.~2004 and
references therein).  The relatively weak
Lyman~$\alpha$ equivalent width of this quasar may be due to dilution by
a beamed continuum.
SDSS~J144231.72+011055.2 ($\alpha_{\rm ox}=-1.37$; $z=4.51$) is a weak-line
quasar discussed in \S4 of Anderson et~al. (2001); the nature of weak-line
quasars remains mysterious. Its relatively strong X-ray emission suggests that
a beamed X-ray continuum component may be present, although it is not a strong
radio source (its integrated FIRST 20~cm flux density of 1.87~mJy indicates
it is only moderately radio loud).  The relatively strong X-ray emission of
SDSS~J144231.72+011055.2 is notably different from that of
SDSS~J153259.96$-$003944.1 (Fan et~al. 1999), the prototype SDSS weak-line
quasar, which is fairly X-ray weak ($\alpha_{\rm ox}<-1.79$; see Table~A1
of Vignali et~al. 2003). There is apparently significant variety among the
weak-line quasar population, even when one considers the time gap
(up to several months in the rest frame) between the optical and X-ray
observations.

\subsection{Close Pairs}

The mechanical constraint
that SDSS spectroscopic fibers must be separated by~55$''$ on a given plate
makes it difficult for the spectroscopic survey to confirm close pairs
of quasars.  In regions that
are covered by more than one plate, however, it is possible
to obtain spectra of both components of a close pair;
there are~121 pairs of quasars in the catalog with angular separation less than
$60''$ (eleven pairs with separations less than 20$''$).
Most of the pairs are chance superpositions, but there are seven pairs that
may be physically associated systems ($\Delta z < 0.02$); they are listed
in Table~5.  Hennawi et al. (2005) identified over 200 physical quasar pairs,
primarily through spectroscopic observations of unconfirmed SDSS
quasar candidates near known SDSS quasars.

\subsection{Morphology}

The images of~2077 of the DR3 quasars are classified as extended by the
SDSS photometric pipeline;~1961~(94\%) have redshifts below one
(there are seven resolved \hbox{$z > 3.0$} quasars).
The majority of the large redshift ``resolved" quasars are probably measurement
errors, but this sample probably also contains a mix of chance superpositions
of quasars and foreground objects or possibly some
small angle separation gravitational lenses (indeed, several lenses
are present in the resolved quasar sample; see Paper~II).

\subsection{Matches with Non-optical Catalogs}

The DR3 Quasar Catalog lists matches in the radio,
X-ray, and infrared bands.  We report radio measurements from the FIRST survey
(Becker, White, \& Helfand~1995).
A total of 3761 catalog objects are FIRST sources (defined by a SDSS-FIRST
positional offset of less than~2.0$''$).  
Extended radio sources may be missed by this matching; to recover at 
least some of these, we separately identify all objects with a greater than
3$\sigma$ detection of FIRST flux at the optical position (1319 sources). 
For these objects as well as those with a FIRST catalog match within 
2$''$, we perform a second FIRST catalog search with 30$''$ matching radius 
to identify possible radio lobes associated with the quasar, finding 
such matches for 891 sources.

Matches with the {\it ROSAT} All-Sky Survey Bright (Voges et al. 1999) and
Faint (Voges et al. 2000) Source Catalogs
were made with a maximum allowed positional offset
of~30$''$.  The SDSS target selection for {\it ROSAT} sources initially
considers SDSS objects that exceed
the~30$''$ catalog matching radius.
The DR3 catalog lists a total of~2672 RASS matches.

The infrared information is provided by the
2MASS All-Sky Data Release Point Source Catalog (Cutri et al.~2003).
The DR3 Quasar Catalog contains the $JHK$ photometric measurements of 
6192 SDSS-2MASS matches (maximum positional offset of~2$''$).

Figures~7a, 7b, and~7c show the distribution of the positional offsets for
the FIRST, RASS, and 2MASS matches, respectively.  The three histograms are
quite similar in shape to the offset distributions found in Paper~II.
The number of chance superpositions between the DR3 Quasar Catalog and the
FIRST and RASS datasets were
estimated by shifting the quasar positions by~{$\pm 200''$.} As was found
in Paper~II, virtually all the FIRST identifications are correct
(an average of two~FIRST ``matches" was found after declination shifting),
while approximately one percent of the {\it ROSAT} matches are
misidentifications
(an average of 20~{\it ROSAT} ``matches" was found after shifting).

\subsection{Redshift Disagreements with Previous Measurements}

The redshifts of~45 quasars in this catalog disagree by more than~0.10
from the values given in the NED database; the information for each
of these objects is given in Table~6 (four of the entries are differences
between the DR1 and DR3 quasar catalogs).  A NED name of ``SDSS" indicates
that the NED entry is taken from a previous SDSS publication.  The
relatively large
number of apparent discrepancies with previous SDSS measurements arises because
the NED redshift for these objects
is frequently the redshift given in the SDSS data release
and not from the quasar catalogs.  For example, the second entry in Table~6
was included in both the EDR and DR1 Quasar Catalogs with the correct redshift,
but the NED value was the redshift reported in the EDR itself.
In every case we believe that the redshifts quoted 
in this catalog are more consistent with the SDSS spectra than are
the NED values.

%
%

\section{Future Work}

The 46,420 quasars were identified from~$\approx$~40\% of the proposed
SDSS survey area.  The progress of the SDSS Quasar Survey can be seen in
Figure~7d, which displays the cumulative number of SDSS quasars
as a function of observing date.  There are yearly ``plateaus" in this figure
which coincide with late summer/fall; at this time of the year the
North Galactic Pole region is unavailable.  The primary spectroscopic
survey of the South Galactic Pole is now complete; observations in this region
now consist
of additional imaging scans (to reach fainter magnitudes; see York et al.~2000)
and a series of specialized spectroscopic programs (e.g., empirical
determination of the effectiveness of the SDSS quasar selection;
Vanden Berk et al.~2005).

Investigations of the quasar luminosity function and the spatial distribution
of quasars based on SDSS data are given in Richards et al.~(2005)
and Yahata et al.~(2005).
We plan to publish the
next edition of the SDSS quasar catalog with the SDSS
Fifth Data Release, which is currently expected to occur in~2006.

\acknowledgments

We thank the referee, Buell Jannuzi, for a number of suggestions that
improved the paper.
This work was supported in part by National Science Foundation grants
AST-0307582 (DPS, DVB), AST-0307384~(XF), and
AST-0307409~(MAS), and by
NASA LTSA grant NAG5-13035 and CXC grant GO3-4117X (WNB, DPS).
XF acknowledges support from an \hbox{Alfred P. Sloan} Fellowship and
a David and Lucile Packard Fellowship in Science and Engineering.
SJ and CS were supported by the U.S. Department of Energy under contract
\hbox{DE-AC02-76CH03000.}

Funding for the creation and distribution of the SDSS Archive
has been provided by the Alfred P. Sloan Foundation, the
Participating Institutions, the National Aeronautics and Space
Administration, the National Science Foundation, the U.S.
Department of Energy, the Japanese Monbukagakusho, and the
Max Planck Society.
The SDSS Web site \hbox{is {\tt http://www.sdss.org/}.}
The SDSS is managed by the Astrophysical Research Consortium
(ARC) for the Participating Institutions.  The Participating
Institutions are
The University of Chicago,
Fermilab,
the Institute for Advanced Study,
the Japan Participation Group,
The Johns Hopkins University,
the Korean Scientist Group,
Los Alamos National Laboratory,
the Max-Planck-Institute for Astronomy (MPIA),
the Max-Planck-Institute for Astrophysics (MPA), 
New Mexico State University, 
University of Pittsburgh,
University of Portsmouth,
Princeton University,
the United States Naval Observatory,
and the University of Washington.

This research has made use of 1)~the NASA/IPAC Extragalactic Database (NED)
which is operated by the Jet Propulsion Laboratory, California Institute
of Technology, under contract with the National Aeronautics and Space
Administration, and 2)~data products from the Two Micron All Sky 
Survey, which is a joint project of the University of
Massachusetts and the Infrared Processing and Analysis Center/California 
Institute of Technology, funded by the National Aeronautics
and Space Administration and the National Science Foundation.

The Hobby-Eberly Telescope (HET) is a joint project of the University of Texas
at Austin,
the Pennsylvania State University,  Stanford University,
Ludwig-Maximillians-Universit\"at M\"unchen, and Georg-August-Universit\"at
G\"ottingen.  The HET is named in honor of its principal benefactors,
William P. Hobby and Robert E. Eberly.  The Marcario Low-Resolution
Spectrograph is named for Mike Marcario of High Lonesome Optics, who
fabricated several optics for the instrument but died before its completion;
it is a joint project of the Hobby-Eberly Telescope partnership and the
Instituto de Astronom\'{\i}a de la Universidad Nacional Aut\'onoma de M\'exico.

\newpage



\begin{figure}
\includegraphics[angle=90,scale=0.70,viewport= 0 0 500 700]{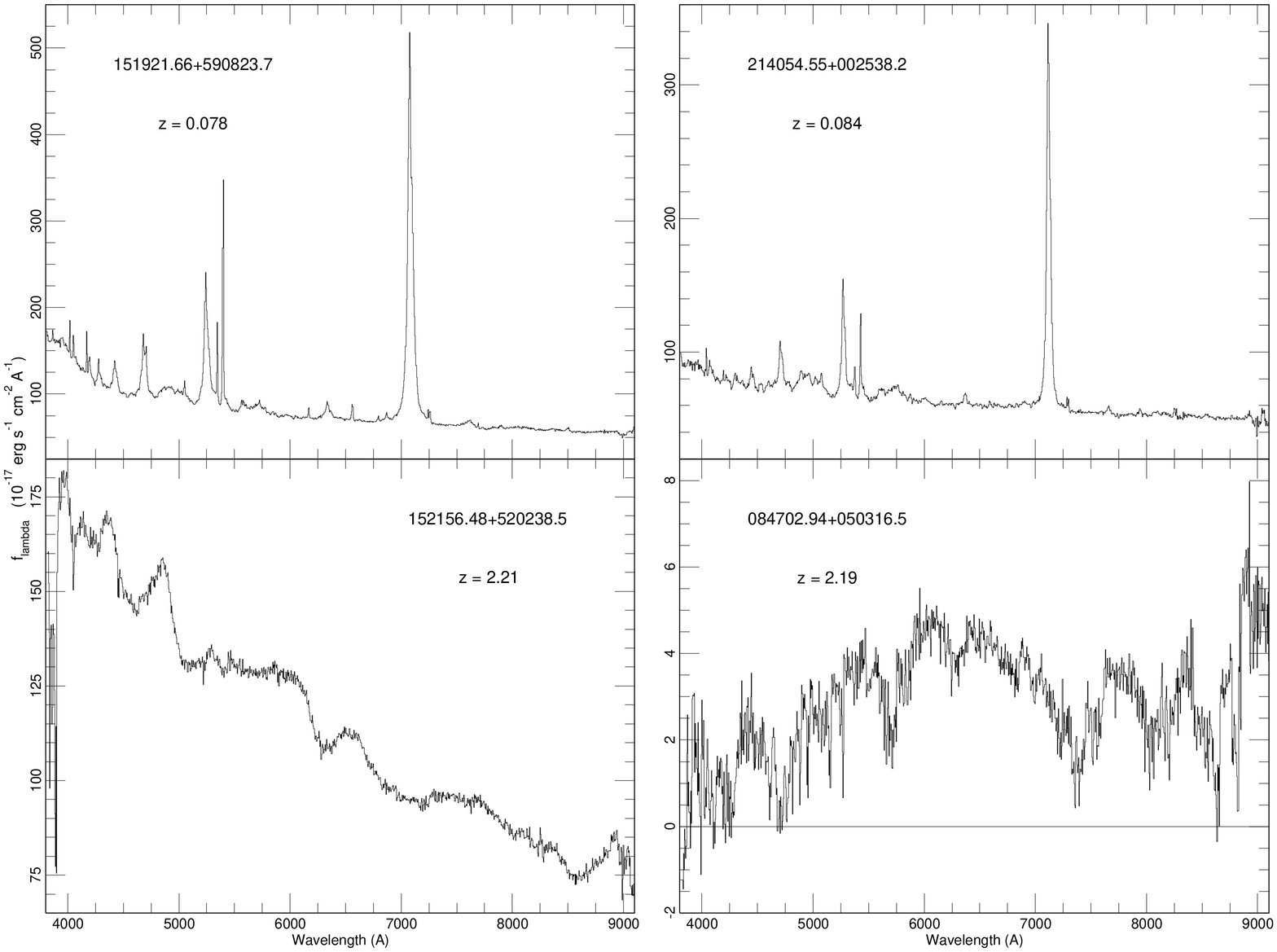}
\caption{
An example of data produced by the SDSS spectrographs.  The spectral
resolution of the data ranges from 1800 to 2100; a dichroic splits the beam
at~6150~\AA .  The data have been rebinned \hbox{to 5 \AA\ pixel$^{-1}$}
for display purposes.  All four of the quasars were discovered by the~SDSS
and are reported here for the first time.
Notes on spectra: \hbox{SDSS J151921.66+590823.7} and
\hbox{SDSS J214054.55+002538.2} are the lowest redshift quasars
among the new objects;
\hbox{SDSS J152156.48+520238.5} is a moderate redshift quasar that is
both bright ($i = 15.44$)
and extremely luminous \hbox{($M_i = -30.2$)};
\hbox{SDSS J084702.94+050516.5}
is an example of a low-ionization BAL with a red continuum and a strong
Mg~II absorption feature at~~$\approx$~8900~\AA .
\label{Figure 1 }
}
\end{figure}

\clearpage

\begin{figure}
\includegraphics[angle=0,scale=0.90]{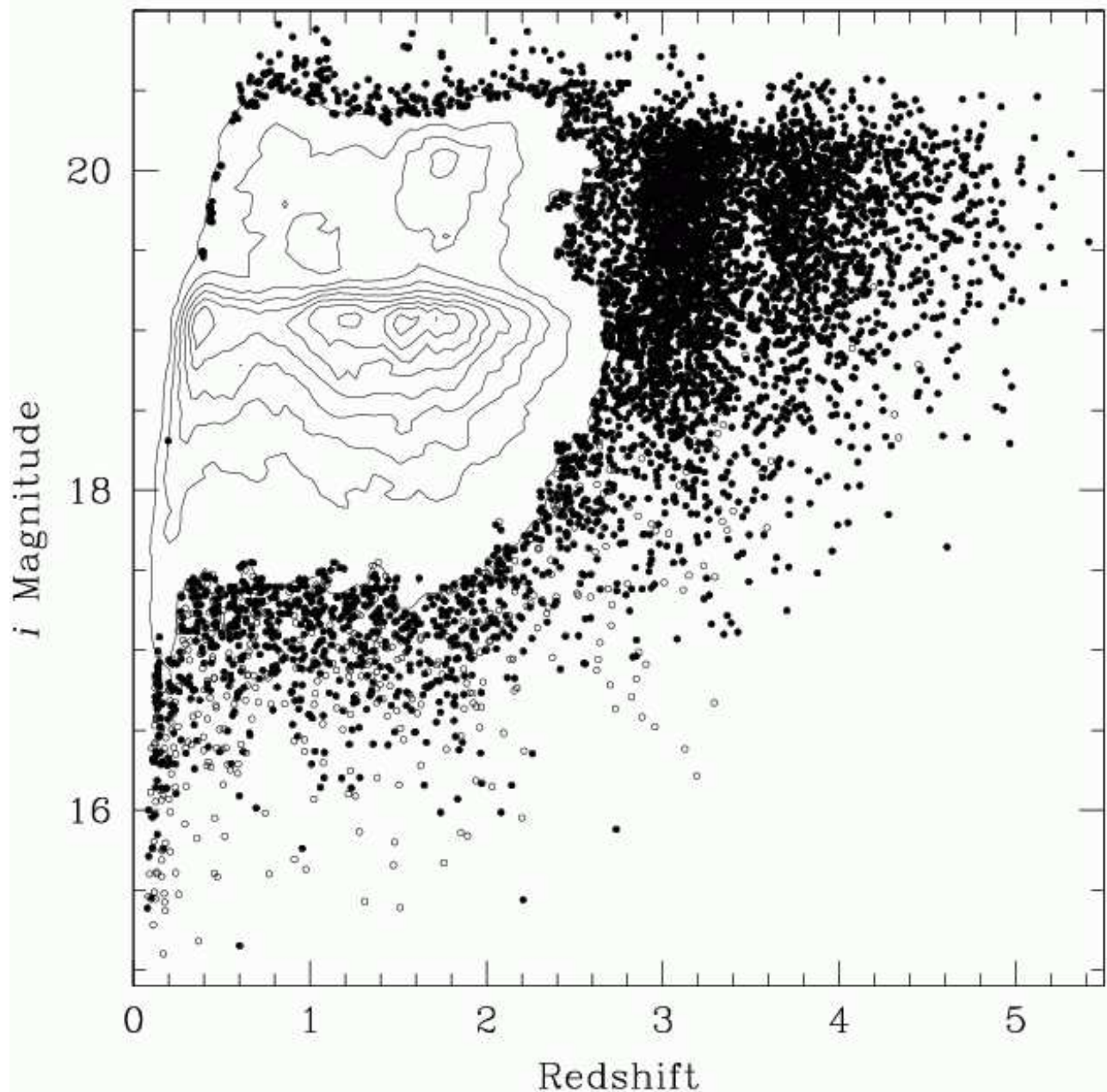}
\figcaption{
The observed~$i$ magnitude as a function of redshift for the~46,420
objects in the catalog.  Open circles indicate quasars in NED that
were recovered but not discovered by
the SDSS.  Five quasars with \hbox{$i > 21$} are not plotted.
The distribution is represented by a set of linear contours when the
density of points in this two-dimensional space exceeds a certain threshold.
\label{Figure 2 }
}
\end{figure}

\clearpage

\begin{figure}
\includegraphics[angle=90,scale=1.10,viewport= 0 0 350 700]{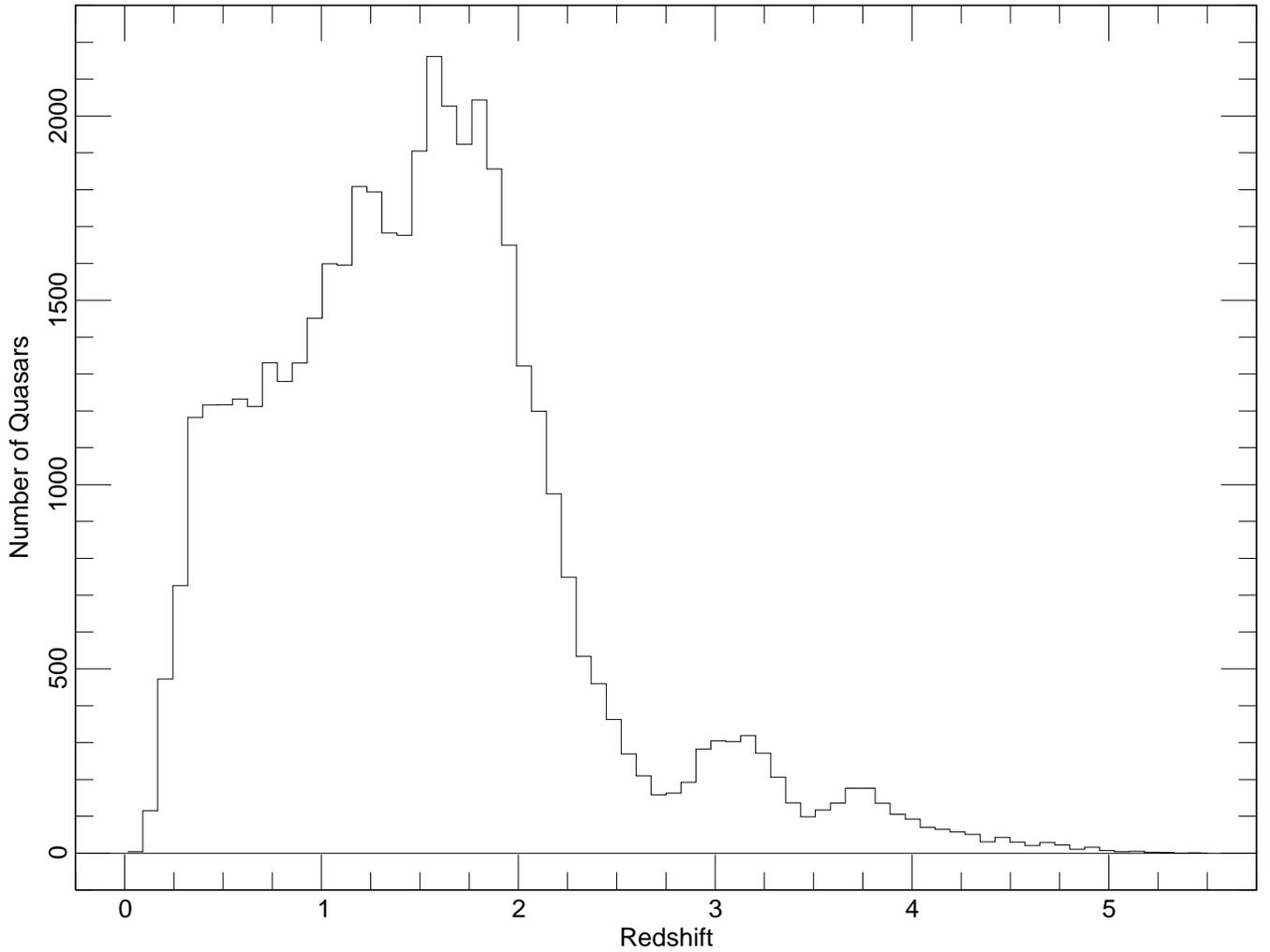}
\figcaption{
The redshift histogram of the catalog quasars.  The smallest redshift is~0.08
and the largest redshift is~5.41; the median redshift of the catalog is~1.47.
The redshift bins have a width of~0.076.  The dips at redshifts of~2.7 and~3.5
are caused by the lower efficiency of the selection algorithm at these
redshifts.
\label{Figure 3 }
}
\end{figure}

\clearpage

\begin{figure}
\includegraphics[angle=90,scale=1.10,viewport= 0 0 350 700]{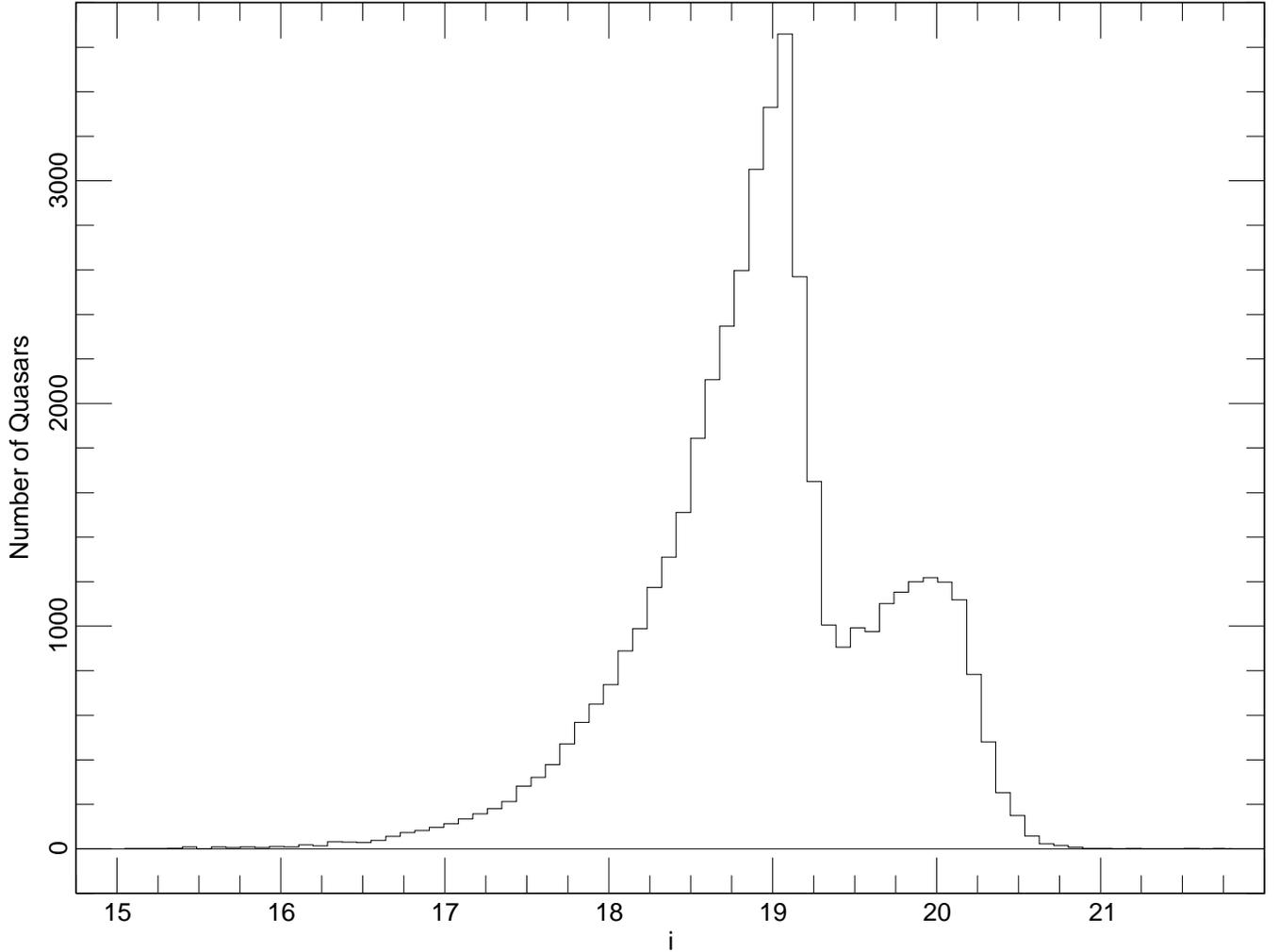}
\figcaption{
The $i$ magnitude (not corrected for Galactic absorption) histogram of the
catalog quasars.  The magnitude bins have a width of~0.089.  The sharp
drop that occurs at magnitudes slightly fainter than~19 is due to the
low-redshift flux limit of the survey.  Quasars fainter than the
\hbox{$i = 20.2$} high-redshift selection limit were found via other
selection algorithms, primarily serendipity.
\label{Figure 4 }
}
\end{figure}

\clearpage

\begin{figure}
\includegraphics[angle=90,scale=1.10,viewport= 0 0 350 700]{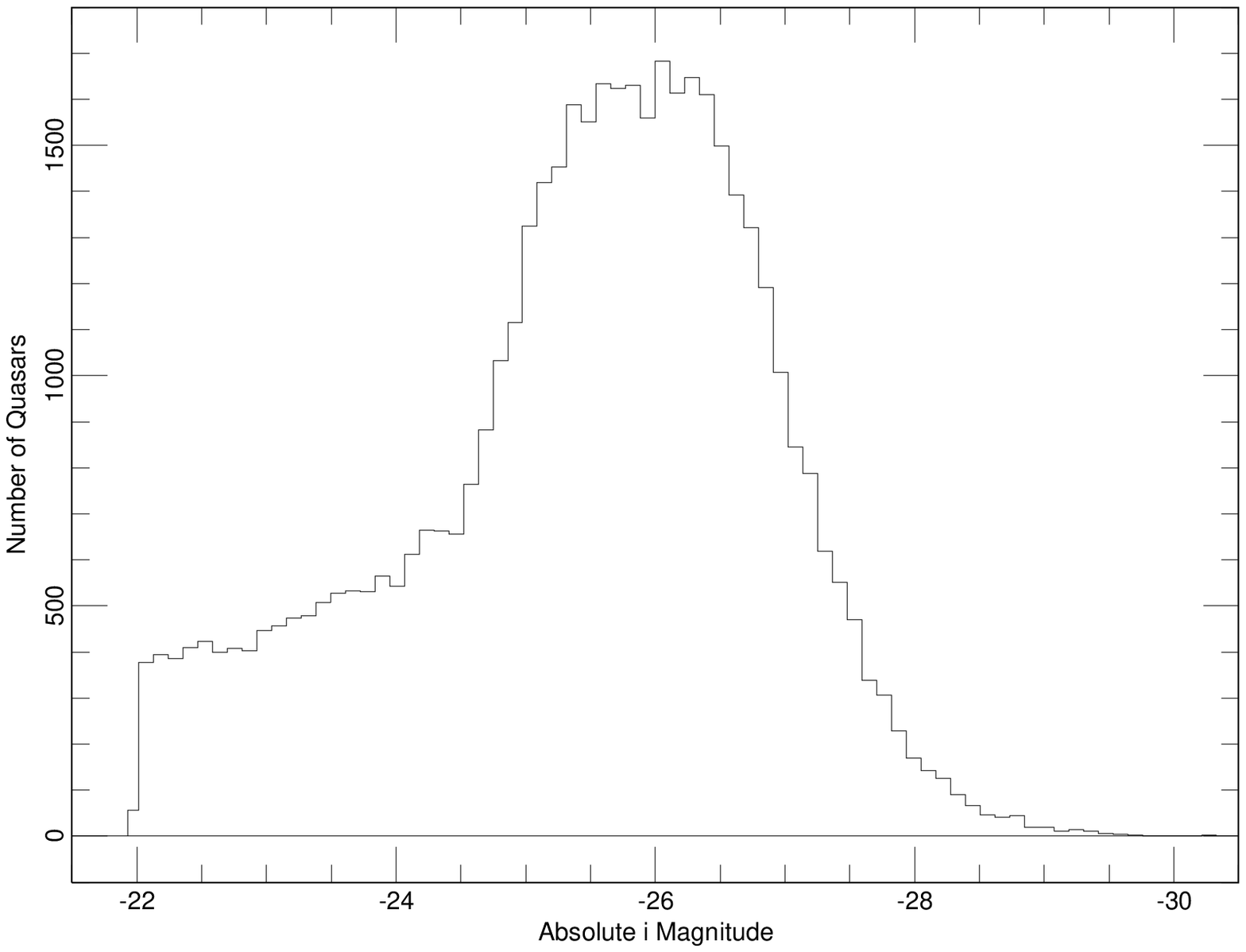}
\figcaption{
The luminosity distribution of the catalog quasars.
The absolute magnitude bins have a width of~0.114.
\label{Figure 5 }
}
\end{figure}

\clearpage

\begin{figure}
\includegraphics[angle=90,scale=0.70,viewport= 0 0 500 700]{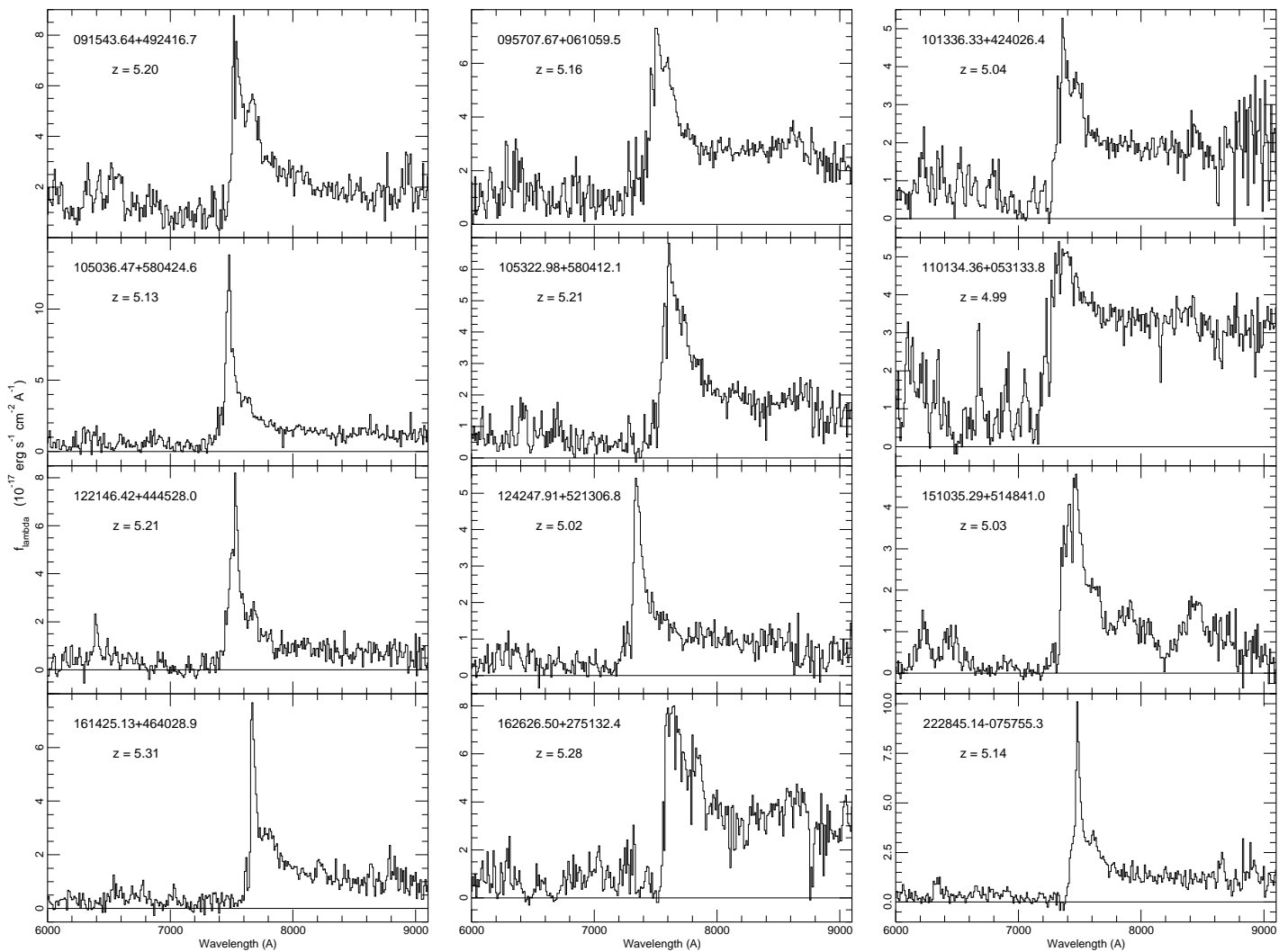}
\figcaption{
SDSS spectra of the 12 new quasars with the highest redshifts ($z \ge 4.99$).
The spectra have been rebinned to \hbox{10 \AA\ pixel$^{-1}$} for display
purposes.  The wavelength region below~6000~\AA\ has been removed because
of the lack of signal in these objects.
\label{Figure 6 }
}
\end{figure}

\clearpage

\begin{figure}
\includegraphics[angle=90,scale=0.70,viewport= 0 0 500 700]{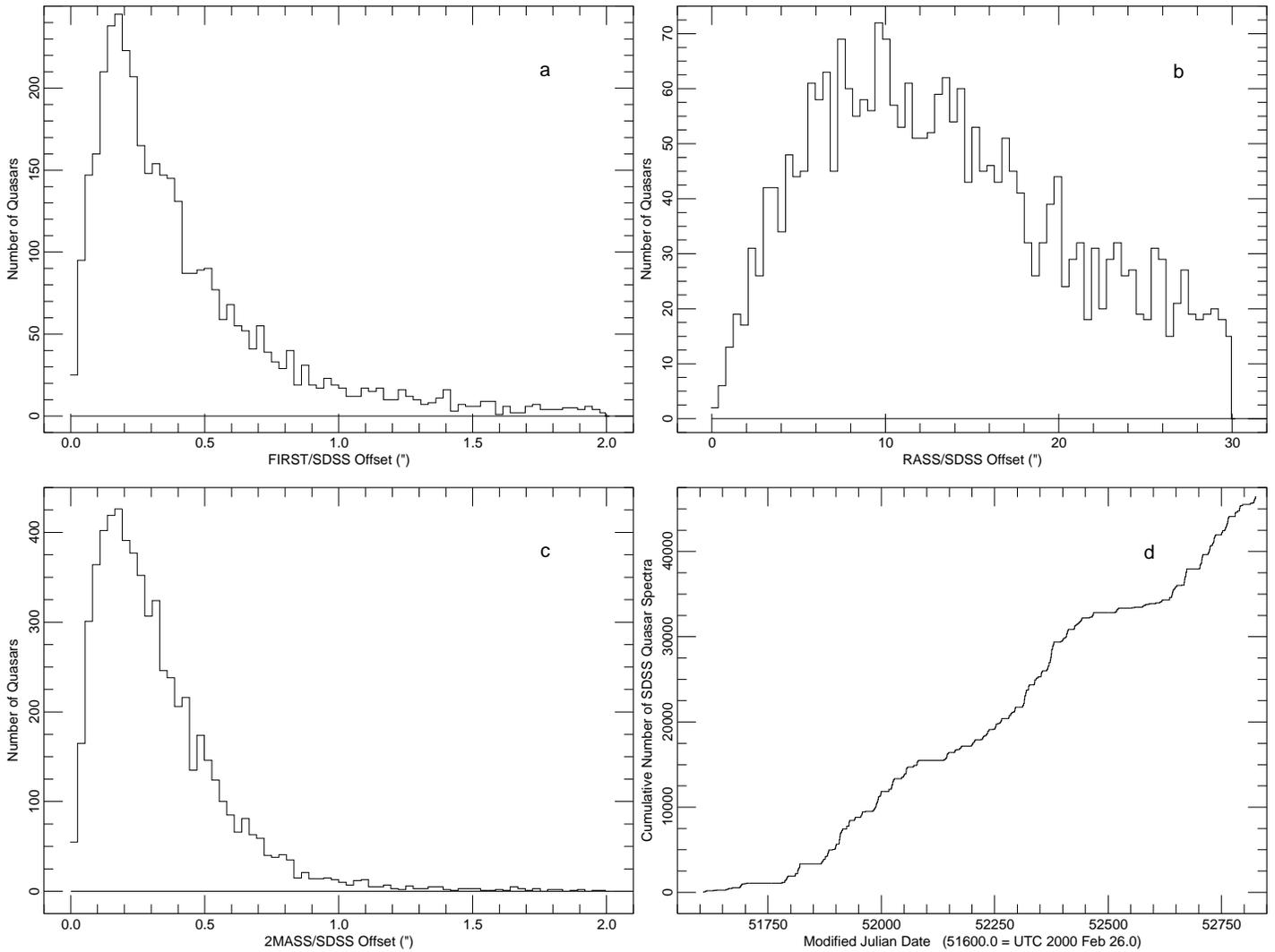}
\figcaption{
a)~Offsets between the 3761 SDSS and FIRST matches; the
matching radius was set to~2.0$''$.
b)~Offsets between the 2672 SDSS and RASS FSC/BSC matches; the
matching radius was set to~30$''$.
c)~Offsets between the 6192 SDSS and 2MASS matches; the
matching radius was set to~2$''$.
d)~The cumulative number of DR3 quasars as a function of time.  The horizonal
axis runs from February 2000 to July~2003.
\label{Figure 7 }
}
\end{figure}

\clearpage


\halign{\hskip 12pt
\hfil # \tabskip=1em plus1em minus1em&
\hfil # \hfil &
# \hfil \cr
\multispan3{\hfil TABLE 1 \hfil} \cr
\noalign{\medskip}
\multispan3{\hfil SDSS DR3 Quasar Catalog Format \hfil} \cr
\noalign{\bigskip\hrule\smallskip\hrule\medskip}
\hfil Column \hfil & \hfil Format \hfil & \hfil Description \hfil \cr
\noalign{\medskip\hrule\bigskip}
   1  &  A18  &   
SDSS DR3 Designation   \ \ \ \ hhmmss.ss+ddmmss.s  \ \ \ (J2000) \cr
   2  &  F11.6  &   Right Ascension in decimal degrees (J2000) \cr
   3  &  F11.6  &   Declination in decimal degrees (J2000) \cr
   4  &  F7.4 &   Redshift \cr
   5  &   F7.3  &    PSF $u$ magnitude
(not corrected for Galactic absorption) \cr
   6  &   F6.3  &    Error in PSF $u$ magnitude \cr
   7  &   F7.3  &    PSF $g$ magnitude
(not corrected for Galactic absorption) \cr
   8  &   F6.3  &    Error in PSF $g$ magnitude \cr
   9  &   F7.3  &    PSF $r$ magnitude 
(not corrected for Galactic absorption) \cr
  10  &   F6.3  &    Error in PSF $r$ magnitude \cr
  11  &   F7.3  &    PSF $i$ magnitude
(not corrected for Galactic absorption) \cr
  12  &   F6.3  &    Error in PSF $i$ magnitude \cr
  13  &   F7.3  &    PSF $z$ magnitude
(not corrected for Galactic absorption) \cr
  14  &   F6.3  &    Error in PSF $z$ magnitude \cr
  15  &   F7.3  &    Galactic absorption in $u$ band \cr
  16  &   F7.3  &    $\log N_H$  (logarithm of Galactic H I column density) \cr
  17  &   F7.3  &    FIRST peak flux density at 20 cm expressed as AB magnitude;
\cr
& & \ \ \ \ \ 0.0 is no detection, $-1.0$ source is not in FIRST area \cr
  18  &   F8.3  &    S/N of FIRST flux density \cr
  19  &   F7.3  &    SDSS-FIRST separation in arc seconds \cr
  20  &   I3    &   $> 3\sigma$ FIRST flux at optical position
but no FIRST counterpart within 2$''$ (0 or 1) \cr
  21  &   I3    &   FIRST source located 2$''$-30$''$ from optical position
(0 or 1) \cr
  22  &   F8.3  &   log RASS full band count rate; $-9.0$ is no detection \cr
  23  &   F7.3  &   S/N of RASS count rate \cr
  24  &   F7.3  &   SDSS-RASS separation in arc seconds \cr
  25  &   F7.3  &   $J$ magnitude (2MASS);
0.0 indicates no 2MASS detection \cr
  26  &   F6.3  &   Error in $J$ magnitude (2MASS) \cr
  27  &   F7.3  &   $H$ magnitude (2MASS);
0.0 indicates no 2MASS detection \cr
  28  &   F6.3  &   Error in $H$ magnitude (2MASS) \cr
  29  &   F7.3  &   $K$ magnitude (2MASS);
0.0 indicates no 2MASS detection \cr
  30  &   F6.3  &   Error in $K$ magnitude (2MASS) \cr
  31  &   F7.3  &   SDSS-2MASS separation in arc seconds \cr
  32  &   F8.3  &   $M_{i}$ ($H_0$ = 70 km s$^{-1}$ Mpc$^{-1}$,
$\Omega_M = 0.3$, $\Omega_{\Lambda} = 0.7$, $\alpha_{\nu} = -0.5$) \cr
\noalign{\medskip\hrule}}

\clearpage

\halign{\hskip 12pt
\hfil # \tabskip=1em plus1em minus1em&
\hfil # \hfil &
# \hfil \cr
\multispan3{\hfil TABLE 1 \hfil} \cr
\noalign{\medskip}
\multispan3{\hfil SDSS DR3 Quasar Catalog Format (Continued) \hfil} \cr
\noalign{\bigskip\hrule\smallskip\hrule\medskip}
\hfil Column \hfil & \hfil Format \hfil & \hfil Description \hfil \cr
\noalign{\medskip\hrule\bigskip}
  33  &   I3  &   Morphology flag \ \ \ 0 = point source \ \ \ 1 = extended \cr
  34  &   I3  &   SDSS SCIENCEPRIMARY flag  (0 or 1) \cr
  35  &   I3  &   SDSS MODE flag (blends, overlapping scans; 1, 2, or 3) \cr
  36  &  I12  &   Target Selection Flag (BEST) \cr
  37  &   I3  &   Low-$z$ Quasar selection flag (0 or 1) \cr
  38  &   I3  &   High-$z$ Quasar selection flag (0 or 1) \cr
  39  &   I3  &   FIRST selection flag (0 or 1) \cr
  40  &   I3  &   {\it ROSAT} selection flag (0 or 1) \cr
  41  &   I3  &   Serendipity selection flag (0 or 1) \cr
  42  &   I3  &   Star selection flag (0 or 1) \cr
  43  &   I3  &   Galaxy selection flag (0 or 1) \cr
  44  &   I6  &   SDSS Imaging Run Number of photometric measurements \cr
  45  &   I6  &   Modified Julian Date of imaging observation \cr
  46  &   I6  &   Modified Julian Date of spectroscopic observation \cr
  47  &   I5  &   Spectroscopic Plate Number \cr
  48  &   I5  &   Spectroscopic Fiber Number \cr
  49  &   I4  &   SDSS Photometric Processing Rerun Number \cr
  50  &   I3  &   SDSS Camera Column Number \cr
  51  &   I5  &   SDSS Frame Number \cr
  52  &   I5  &   SDSS Object Number \cr
  53  &   I4  &   SDSS Chunk Number \cr
  54  &  I12  &   Target Selection Flag (TARGET) \cr
  55  &   F7.3  &    TARGET PSF $u$ magnitude
(not corrected for Galactic absorption) \cr
  56  &   F6.3  &    TARGET Error in PSF $u$ magnitude \cr
  57  &   F7.3  &    TARGET PSF $g$ magnitude
(not corrected for Galactic absorption) \cr
  58  &   F6.3  &    TARGET Error in PSF $g$ magnitude \cr
  59  &   F7.3  &    TARGET PSF $r$ magnitude 
(not corrected for Galactic absorption) \cr
  60  &   F6.3  &    TARGET Error in PSF $r$ magnitude \cr
  61  &   F7.3  &    TARGET PSF $i$ magnitude
(not corrected for Galactic absorption) \cr
  62  &   F6.3  &    TARGET Error in PSF $i$ magnitude \cr
  63  &   F7.3  &    TARGET PSF $z$ magnitude
(not corrected for Galactic absorption) \cr
  64  &   F6.3  &    TARGET Error in PSF $z$ magnitude \cr
  65  &   1X, A25 &   Object Name for previously known quasars \cr
 & & \ \ \ ``SDSS" designates previously published SDSS object \cr
\noalign{\medskip\hrule}}

\clearpage

\begin{deluxetable}{rrrrrrrrrrrrrr}
\tablecolumns{14}
\tabletypesize{\scriptsize}
\rotate
\tablewidth{8in}
\tablenum{2}
\tablecaption{The SDSS Quasar Catalog III\tablenotemark{a}}
\tablehead{
  \colhead{Object (SDSS J)} &
  \colhead{R.A.} &
  \colhead{Dec.} &
  \colhead{Redshift} &
  \multicolumn{2}{c}{u} &
  \multicolumn{2}{c}{g} &
  \multicolumn{2}{c}{r} &
  \multicolumn{2}{c}{i} &
  \multicolumn{2}{c}{z}
}
\startdata
000009.26$+151754.5$  & 0.038605 & 15.298476 & 1.1986 &
19.921 & 0.042 & 19.811 & 0.036 & 19.386 & 0.017 & 19.165 & 0.023 & 19.323 &
 0.069 \\
000009.38$+135618.4$  & 0.039088 & 13.938447 & 2.2400 &
19.218 & 0.026 & 18.893 & 0.022 & 18.445 & 0.018 & 18.331 & 0.024 & 18.110 &
 0.033 \\
000009.42$-102751.9$  & 0.039269 & -10.464428 & 1.8442 &
19.249 & 0.036 & 19.029 & 0.027 & 18.980 & 0.021 & 18.791 & 0.018 & 18.751 &
 0.047 \\
000011.41$+145545.6$  & 0.047547 & 14.929353 & 0.4596 &
19.637 & 0.030 & 19.466 & 0.024 & 19.362 & 0.022 & 19.193 & 0.025 & 19.005 &
 0.047 \\
000011.96$+000225.3$  & 0.049842  & 0.040372 & 0.4790 &
18.237 & 0.028 & 17.971 & 0.020 & 18.025 & 0.019 & 17.956 & 0.014 & 17.911 &
 0.029 \\
\enddata
\tablenotetext{a}{Table 2 is presented in its entirety in the electronic
  edition of the Astronomical Journal.  A portion is shown here for guidance
  regarding its form and content.  The full catalog contains 65 columns
  of information on 46,420 quasars.}
\end{deluxetable}

\clearpage

\halign{\hskip 12pt
# \hfil \tabskip=1em plus1em minus1em&
\hfil # &
\hfil # \cr
\multispan3{\hfil TABLE 3 \hfil} \cr
\noalign{\medskip}
\multispan3{\hfil Spectroscopic Target Selection (BEST) \hfil} \cr
\noalign{\bigskip\hrule\smallskip\hrule\medskip}
&&\hfil Sole \hfil \cr
\hfil Class \hfil & \hfil Selected \hfil & \hfil Selection \hfil \cr
\noalign{\medskip\hrule\bigskip}
Low-$z$ & 28831 & 8738 \cr
High-$z$ & 9852 & 2286 \cr
FIRST  & 2173 & 114 \cr
{\it ROSAT}  & 3146 & 366 \cr
Serendipity & 25565 & 9703 \cr
Star & 495 & 108 \cr
Galaxy & 417 & 57 \cr
\noalign{\medskip\hrule}}

\clearpage

\halign{\hskip 12pt
$#$ \hfil \tabskip=1em plus1em minus1em&
# \hfil &
$#$ \hfil &
# \hfil \cr
\multispan4{\hfil TABLE 4 \hfil} \cr
\noalign{\medskip}
\multispan4{\hfil DR1 Quasars not in DR3 Quasar Catalog \hfil} \cr
\noalign{\bigskip\hrule\smallskip\hrule\medskip}
\hfil $DR1  SDSS J$ \hfil & \hfil Comment \hfil &
 \hfil $DR1  SDSS J$ \hfil & \hfil Comment \hfil \cr
\noalign{\medskip\hrule\bigskip}
014349.15+002128.3   & Visual Exam &
103951.49+643004.1   & $v < 1000$ km s$^{-1}$ \cr
021047.00-100153.0   & $v < 1000$ km s$^{-1}$ &
104718.29+000718.4   & Visual Exam \cr
022011.27-000636.2   & $M_i > -22.0$ &
114245.45+680236.5   & $v < 1000$ km s$^{-1}$ \cr
022119.84+005628.5   & $v < 1000$ km s$^{-1}$ &
115404.55-001009.8   & Visual Exam \cr
024921.52-080957.1   & Unmapped Fiber &
125751.28-004350.6   & $v < 1000$ km s$^{-1}$ \cr
025448.47-071735.3   & Visual Exam &
132444.10-021746.5   & Visual Exam \cr
033606.70-000754.7   & Visual Exam &
141515.46+654615.1   & $M_i > -22.0$ \cr
080154.25+441233.9   & $v < 1000$ km s$^{-1}$ &
142505.59+035336.2   & Visual Exam \cr
083223.22+491320.9   & Visual Exam &
144248.53+001540.4   & Visual Exam \cr
091040.26-004226.3   & Visual Exam &
144454.60+004224.5   & $v < 1000$ km s$^{-1}$ \cr
091507.24+545825.2   & $M_i > -22.0$ &
151304.44+021602.3   & 151304.35+021603.8 \cr
093233.53+553620.8   & Unmapped Fiber &
151320.82+024128.5   & 151320.61+024130.1 \cr
093717.70+011836.6   & $v < 1000$ km s$^{-1}$ &
152200.77+035017.0   & $v < 1000$ km s$^{-1}$ \cr
093859.26+020924.3   & Visual Exam &
153133.69+010136.9   & Visual Exam \cr
094542.23+575747.6   & Visual Exam &
155936.46+025601.5   & $v < 1000$ km s$^{-1}$ \cr
094620.21+010452.0   & Visual Exam &
163353.22+431111.0   & Visual Exam \cr
101317.22-011056.2   & $M_i > -22.0$ &
164444.93+423304.5   & $M_i > -22.0$ \cr
101355.20+002850.2   & $M_i > -22.0$ &
212723.01-073726.8   & Unmapped Fiber \cr
102423.28-004800.5   & $M_i > -22.0$ &
214828.94-064145.8   & $v < 1000$ km s$^{-1}$ \cr
102746.04+003204.9   & $v < 1000$ km s$^{-1}$ &
232137.82+011128.9   & $v < 1000$ km s$^{-1}$ \cr
102757.09+005406.9   & $M_i > -22.0$ &
235818.87-000919.4   & $v < 1000$ km s$^{-1}$ \cr
103458.60-004553.8   & $M_i > -22.0$ \cr
\noalign{\medskip\hrule}}

\clearpage

\halign{\hskip 12pt
\hfil # \hfil \tabskip=1em plus1em minus1em&
\hfil # \hfil &
\hfil # \hfil &
\hfil # \hfil &
\hfil # \cr
\multispan5{\hfil TABLE 5 \hfil} \cr
\noalign{\medskip}
\multispan5{\hfil Candidate Binary Quasars \hfil} \cr
\noalign{\bigskip\hrule\smallskip\hrule\medskip}
\hfil Quasar 1 \hfil & \hfil Quasar 2 \hfil &
 \hfil $z_1$ \hfil & \hfil $z_2$ \hfil & \hfil $\Delta \theta$ ($''$) \hfil \cr
\noalign{\medskip\hrule\bigskip}
025959.68+004813.6 & 030000.57+004828.0 & 0.892 & 0.900 & 19.6 \cr
074336.85+205512.0 & 074337.28+205437.1 & 1.570 & 1.565 & 35.5 \cr
085625.63+511137.0 & 085626.71+511117.8 & 0.543 & 0.543 & 21.8 \cr
090923.12+000204.0 & 090924.01+000211.0 & 1.884 & 1.865 & 15.0 \cr
095556.37+061642.4 & 095559.02+061701.8 & 1.278 & 1.273 & 44.0 \cr
110357.71+031808.2 & 110401.48+031817.5 & 1.941 & 1.923 & 57.3 \cr
121840.47+501543.4 & 121841.00+501535.8 & 1.457 & 1.455 &  9.1 \cr
\noalign{\medskip\hrule}}

\clearpage
{\footnotesize

\halign{\hskip 12pt
$#$ \hfil \tabskip=1em plus1em minus1em&
\hfil # \hfil &
\hfil $#$ \hfil &
# \hfil \cr
\multispan4{\hfil TABLE 6 \hfil} \cr
\noalign{\smallskip}
\multispan4{\hfil DR3/NED Discrepant Redshifts \hfil} \cr
\noalign{\smallskip\hrule\smallskip\hrule\smallskip}
\hfil $SDSS J$ \hfil & \hfil $z_{\rm DR3}$ \hfil &
\hfil  z_{\rm NED} - z_{\rm DR3} & NED Object Name$^a$ \cr
\noalign{\smallskip\hrule\smallskip}
 002411.65-004348.0 &   1.794 & -1.023 &   LBQS 0021-0100            \cr
 012428.09-001118.4 &   1.728 & -1.213 &   SDSS \cr
 021249.59+003448.7 &   1.222 & -1.058 &   SDSS \cr
 023044.81-004658.0 &   0.916 & +0.908 &   SDSS \cr
 033305.32-053708.9 &   4.218 & -0.128 &   SDSS \cr
 083148.29+463650.0 &   0.830 & +2.330 &   SDSS \cr
 084035.41+412645.6 &   0.866 & +0.284 &   2MASSi J0840353+412645    \cr
 091446.26+400304.6 &   0.322 & +0.614 &   NGC 2782 U1               \cr
 092004.31+591732.7 &   1.286 & -0.724 &   SBS 0916+595              \cr
 093052.25+003458.8 &   1.771 & -1.266 &   [HB89] 0928+008           \cr
 094443.08+580953.2 &   0.561 & +0.146 &   SBS 0941+583              \cr
 095227.30+504850.6 &   1.091 & +0.455 &   SBS 0949+510              \cr
 095723.69+011458.7 &   2.466 & -1.543 &   SDSS \cr
 101104.39+005724.0 &   0.683 & -0.417 &   SDSS \cr
 101139.85+004039.5 &   1.712 & +0.606 &   SDSS \cr
 101211.63+003719.4 &   1.631 & -0.891 &   SDSS \cr
 103351.42+605107.3 &   1.401 & -1.065 &   87GB 103034.2+610640      \cr
 103506.01+565257.9 &   1.855 & -1.278 &   87GB 103154.7+570825      \cr
 103626.33+045436.4 &   1.049 & -0.667 &   [CC91] 103350.4+051010    \cr
 104207.56+501321.9 &   1.265 & -0.915 &   4C +50.31                 \cr
 104612.40+033624.9 &   2.112 & -0.632 &   [CC91] 104337.3+035213    \cr
 111922.35+604851.3 &   2.014 & -1.716 &   SBS 1116+610              \cr
 113049.37-025048.2 &   1.408 & -0.670 &   2QZ J113049.3-025049      \cr
 114106.59-014107.5 &   1.422 & -0.152 &   2QZ J114106.5-014108      \cr
 114534.12+010308.0 &   1.076 & -0.674 &   SDSS \cr
 120015.35+000553.1 &   1.650 & -1.281 &   SDSS \cr
 124129.57+602041.2 &   2.069 & -0.612 &   2MASSi J1241296+602041    \cr
 125339.67+510203.7 &   1.565 & -0.810 &   SBS 1251+513              \cr
 130907.23+560326.6 &   2.155 & -2.138 &   SBS 1307+563              \cr
 133121.81+000248.4 &   3.218 & -2.352 &   SDSS \cr
 134050.48-013449.2 &   0.744 & +0.726 &   UM 600                    \cr
 134809.71-013847.1 &   0.897 & +1.537 &   2QZ J134809.7-013848      \cr
 140846.46+552336.0 &   2.022 & -1.362 &   SBS 1407+556              \cr
 140848.81+650528.0 &   1.937 & -0.927 &   SDSS \cr
 141454.73+552802.5 &   2.031 & -1.360 &   SBS 1413+556              \cr
 143233.96-012145.4 &   0.521 & +1.972 &   2QZ J143233.9-012145      \cr
 143601.58+002042.0 &   1.787 & -0.522 &   SDSS \cr
 151307.26-000559.3 &   1.022 & +0.838 &   SDSS \cr
 152529.92-001537.4 &   0.758 & +1.442 &   SDSS \cr
 152715.82+491833.4 &   2.270 & -0.668 &   SBS 1525+494              \cr
 161351.34+374258.7 &   0.808 & +0.822 &   [HB89] 1612+378           \cr
 162111.27+374604.9 &   1.273 & -0.933 &   4C +37.46                 \cr
 163647.07+385501.6 &   1.100 & +0.484 &   [CCS88] 163503.7+390101   \cr
 163815.04+392731.9 &   0.573 & +0.717 &   [CCS88] 163633.0+393322   \cr
 171930.24+584804.7 &   2.081 & -1.376 &   SDSS \cr
\noalign{\smallskip\hrule}}

\bigskip
$^a$ NED references:
[CC91]~--~Clowes \& Campusano~(1991); 
[CCS88]~--~Crampton et al.~(1988);
[HB89]~--~Hewett \& Burbidge~(1989).

}
\clearpage

\end{document}